\documentclass[a4paper,10pt,twoside]{just}

\usepackage{multicol}
\usepackage{graphicx}
\usepackage{booktabs}
\usepackage{amssymb,bm,mathrsfs,bbm,amscd}
\usepackage[tbtags]{amsmath}
\usepackage{lastpage}
\usepackage{CJK}

\usepackage{xcolor} 
\usepackage{setspace} 
\usepackage{multirow}
\usepackage{multicol}  
\usepackage{hyperref}
\usepackage{comment}
\usepackage{threeparttable} 

\newsavebox{\tablebox}

\def\BES {\text{BES{\ }\uppercase\expandafter{\romannumeral3}}}
\def\BTWO {\text{Belle{\ }\uppercase\expandafter{\romannumeral2}}}
\def\disc {\textcolor{red}{$\bigstar$}}
\def\evid {\textcolor{magenta}{$\heartsuit$}}
\def\meas {\textcolor{blue}{$\checkmark$}}
\def\ddbar {D^0\text{-}\bar{D^0}}

\begin{document}
\begin{CJK*}{GBK}{song}

\fancyhead[c]{\small  10th International Workshop on $e^+e^-$ collisions from $\phi$ to $\psi$ (PhiPsi15)}
 \fancyfoot[C]{\small PhiPsi15-\thepage}

\footnotetext[0]{Received 14 Sep. 2015}

\title{Charm studies at Belle experiment\thanks{Supported by National Natural Science
Foundation of China (No.11475164) }}

\author{%
      Longke Li$^{1)}$\email{lilongke@mail.ustc.edu.cn}
%\quad Ziping Zhang$^{1} $ %\email{zpz@ustc.edu.cn} 
%\quad Wenbiao Yan$^{1} $ %\email{wenbiao@ustc.edu.cn} 
 (for the Belle Collaboration)
}
\maketitle

\address{%
%$^1$
 Department of Modern Physics,University of Science and Technology of China, Hefei 230026, China\\
%$^2$ {\bf Example}: Department of  Physics,University of Science and Technology of China, Hefei 230026, China\\
}

\begin{abstract}
In this proceeding, we give a summary of the new published or preliminary experimental results on charm studies 
at Belle experiment at KEKB. It mainly includes three parts:
(1) $\ddbar$ mixing and CP violation. Some decay channels give the new measurement results with more precise;
(2) search for $D^0$ rare decay with most restrictive upper limit;
(3) some new results for charmed baryon studies, such as doubly charmed baryons, charmed strange baryon and so on.
\end{abstract}



\begin{multicols}{2}
\section{Introduction}
Belle experiment at KEKB, an asymmetric-energy $e^+e^-$ collider, located at Tsukuba in Japan, and has the world highest peak luminosity $2.1\times10^{34}$ $cm^{-2}s^{-1}$, collected most data at or near the $\Upsilon(nS)$ resonances (n=1, 2, 3, 4, 5) with integrated luminosity about 1 $ab^{-1}$. Belle detector has good momentum resolution and vertex resolution and can separate kaon and pion mesons up to $\sim3.5$ GeV/$c$.  A detailed description of Belle detector can be found in Ref. \cite{Belle}. 

In this proceeding, we give a summary of the new published or preliminary experimental results on charm studies at Belle since last PhiPsi workshop in 2013. It mainly includes three parts: $\ddbar$ mixing and charge-conjugation and parity (CP) violation, $D^0$ rare decay and charmed baryons spectroscopy.

\section{$\ddbar$ mixing and CP violation}
Since $\ddbar$ mixing, as the only up-type quark meson mixing, has already observed with the confidence level of more than $5\sigma$ in single decay channel\cite{obs_LHCb, obs_CDF, lab1} in recent years, all open-flavored neutral meson mixing phenomena, originated from the difference between the flavor and mass eigenstates of the meson-antimeson system, are well established. 
The mixing is described by two parameters, $x=\Delta m/\Gamma$ and $y=\Delta\Gamma/(2\Gamma)$, where $\Delta m$ and $\Delta\Gamma$ are the mass and width differences between the two mass eigenstates and $\Gamma$ is the average decay width of the mass eigenstates.
The mixing parameters $x$ and $y$ are difficult to calculate. The Standard Model (SM) predicts that $\ddbar$ mixing can occur via short distance effects and long distance effects and is strongly suppressed to $\sim1\%$ in charm system.

There are three types of CP violation (CPV) according to their different sources: (1) in the decay (direct CPV): $|\bar{A}_{\bar{f}}/A_{f}|\neq1$; (2) in the mixing (indirect): $r_m=|q/p|\neq1$; (3) in the interference between mixing and decay: $\arg(q/p)\neq 0$. Here we defined the amplitude of $D^0$ decays: $\langle f|\mathcal{H}|D^0\rangle=A_{f}$, $\langle \bar{f}|\mathcal{H}|\bar{D^0}\rangle=\bar{A}_{\bar{f}}$. The status of all experiments referred to HFAG\cite{HFAG} has shown in Table \ref{tab1}. We can see only one single decay channel has given the observation measurement for $\ddbar$ mixing and two or three decay channels given the evidence for $\ddbar$ mixing and CPV. We need to give the observation or evidence in more channels. 

\end{multicols}
\begin{table}
\begin{center}
\tabcaption{\label{tab1} Status of all experiments referred to HFAG\cite{HFAG}.}
\footnotesize
\begin{tabular*}{170mm}{@{\extracolsep{\fill}}l|l|llllll} 
\toprule \hline %\hline
\bf{Decay Type}		& \bf{Final State} 		& \bf{LHCb}  &  \bf{Belle}  & \bf{BABAR} & \bf{CDF}  	& \bf{CLEO}  	& \bf{$\BES$}	 		\\ \hline \hline
DCS 2-body (WS)	& $K^+\pi^-$ 	     	& \href{http://journals.aps.org/prl/abstract/10.1103/PhysRevLett.111.251801}{\disc}    & \href{http://journals.aps.org/prl/abstract/10.1103/PhysRevLett.112.111801}{\disc}   & \href{http://journals.aps.org/prl/abstract/10.1103/PhysRevLett.84.5038}{\evid} 	  & \href{http://journals.aps.org/prl/abstract/10.1103/PhysRevLett.111.231802}{\disc}    	&  \href{http://journals.aps.org/prl/abstract/10.1103/PhysRevLett.84.5038}{\meas} 	& \href{http://www.sciencedirect.com/science/article/pii/S0370269314003815}{\meas\text{${}_{\delta^{K\pi}}$}} 	\\ \hline 
CP-eigenstates		& $K^+K^-$, $\pi^+\pi^-$     	& \href{http://journals.aps.org/prl/abstract/10.1103/PhysRevLett.108.111602}{\evid \text{${}_{A_{CP}}$}}  & \href{http://arxiv.org/abs/1509.08266}{\evid}   & \href{http://journals.aps.org/prd/abstract/10.1103/PhysRevD.87.012004}{\evid}    & \href{http://journals.aps.org/prd/abstract/10.1103/PhysRevD.65.092001}{\meas\text{${}_{A_{CP}}$}} 	&  \href{http://journals.aps.org/prd/abstract/10.1103/PhysRevD.65.092001}{\meas}	&  				\\ \hline 
DCS 3-body (WS)	& $K^+\pi^-\pi^0$  	&            & \href{http://journals.aps.org/prl/abstract/10.1103/PhysRevLett.95.231801}{\meas\text{${}_{A_{CP}}$}}    & \href{http://journals.aps.org/prl/abstract/10.1103/PhysRevLett.103.211801}{\evid}   &         &   \href{http://journals.aps.org/prl/abstract/10.1103/PhysRevLett.87.071802}{\meas\text{${}_{A_{CP}}$}}	&            	\\ \hline 
SCS 3-body 		& $K_S^0K^{\pm}\pi^{\mp}$  
							&  	      &       &	  &           	& \href{http://journals.aps.org/prd/abstract/10.1103/PhysRevD.85.092016}{\meas\text{${}_{\delta^{K_{S}^{0}K\pi}}$}}    & 		     		\\\hline 
Semileptonic decay	& $K^+\ell^-\nu_\ell$	&            & \href{http://journals.aps.org/prd/abstract/10.1103/PhysRevD.77.112003}{\meas}  & \href{http://journals.aps.org/prd/abstract/10.1103/PhysRevD.76.014018}{\meas} &          	& \href{http://journals.aps.org/prd/abstract/10.1103/PhysRevD.71.077101}{\meas}    & 		  		\\ \hline 
\multirow{2}{*}{Self-conjugated 3-body}
				& $K_S^0\pi^+\pi^-$ 	&            & \href{http://journals.aps.org/prd/abstract/10.1103/PhysRevD.89.091103}{\meas}  & \href{http://journals.aps.org/prl/abstract/10.1103/PhysRevLett.105.081803}{\meas}  & \href{http://journals.aps.org/prd/abstract/10.1103/PhysRevD.86.032007}{\meas\text{${}_{A_{CP}}$}}  & \href{http://journals.aps.org/prd/abstract/10.1103/PhysRevD.72.012001}{\meas}    & 		    		\\ %\cline{2-8}
				& $K_S^0K^+K^-$   	&            & \href{http://journals.aps.org/prd/abstract/10.1103/PhysRevD.80.052006}{\meas${}^{(a)}$}  & \href{http://journals.aps.org/prl/abstract/10.1103/PhysRevLett.105.081803}{\meas} &       	&  		&       	\\ \hline
\multirow{2}{*}{Self-conjugated SCS 3-body}
				& $\pi^+\pi^-\pi^0$	&	      & \href{http://www.sciencedirect.com/science/article/pii/S0370269308002517}{\meas\text{${}_{A_{CP}}$}}  &  \href{http://journals.aps.org/prd/abstract/10.1103/PhysRevD.78.051102}{\meas\text{${}_{A_{CP}}$}} &		&		&				\\ %\cline{2-8}
				& $K^+K^-\pi^0$	&	      &             &	 \href{http://journals.aps.org/prd/abstract/10.1103/PhysRevD.78.051102}{\meas\text{${}_{A_{CP}}$}} &		&		&				\\\hline 
\multirow{3}{*}{multi-body ($n\ge4$)}
			& $\pi^+\pi^-\pi^+\pi^-$&  \href{http://www.sciencedirect.com/science/article/pii/S0370269313007284}{\meas\text{${}_{A_{CP}}$}} &  &  &          	&  		&        			\\ %\cline{2-8}
			& $K^+\pi^-\pi^+\pi^-$&            & \href{http://journals.aps.org/prl/abstract/10.1103/PhysRevLett.95.231801}{\meas\text{${}_{A_{CP}}$}}  & \href{http://arxiv.org/abs/hep-ex/0607090}{\meas} &          	&  		&        			\\ %\cline{2-8}
			& $K^+K^-\pi^+\pi^-$	 & \href{http://www.sciencedirect.com/science/article/pii/S0370269313007284}{\meas\text{${}_{A_{CP}}$}}\href{http://link.springer.com/article/10.1007\%2FJHEP10\%282014\%29005}{\text{${}_{A_{T}}$}}  &   & \href{http://journals.aps.org/prd/abstract/10.1103/PhysRevD.81.111103}{\meas\text{${}_{A_{T}}$}}  &          	&  \href{http://journals.aps.org/prd/abstract/10.1103/PhysRevD.85.122002}{\meas\text{${}_{A_{CP}}$}}  		&        			\\ \hline 
\multicolumn{2}{c|}{ $\psi(3770)\to D^0\bar{D^0}$ via quantum correlations} 
 			 				& 	     & 		   & 		 & 	      	& \href{http://journals.aps.org/prd/abstract/10.1103/PhysRevD.86.112001}{\meas\text{${}_{\delta^{K\pi}}$}} 	&\href{http://www.sciencedirect.com/science/article/pii/S0370269315002518}{\meas}	     		\\
\hline %\hline
\bottomrule
\end{tabular*}
\begin{threeparttable}
\begin{tablenotes}
  \item[PS:] \textcolor{red}{$\bigstar$} for observation ($>5\sigma$); \textcolor{magenta}{$\heartsuit$} for evidence ($>3\sigma$); \textcolor{blue}{\checkmark} for measurement; more analyses on going are not included.  \\
  The published references are linked under their corresponding signs.
  \item[(a)] Belle exactly measured $y_{CP}$ in $D^0\to K_S^0\phi$ in Phys. Rev. D {\bf 80} 052006 (2009). 
\end{tablenotes}
\end{threeparttable}
\end{center}
\end{table}

\begin{multicols}{2}

\subsection{Wrong-sign decay $D^0\to K^+\pi^-$\cite{lab1}}
Belle gave the first observation of $\ddbar$ mixing for an $e^+e^-$ collision experiment by measuring the time-dependent ratio of $D^0\to K^+\pi^-$ wrong-sign (WS) decay, to $D^0\to K^-\pi^+$ right-sign (RS) decay rates using a data sample of integrated luminosity 976 $fb^{-1}$. We tag the RS and WS decays through the decay chain $D^{*\pm}\to D^0(\to K^{\mp}\pi^{\pm})\pi_{s}^+$  by comparing the charge of the $\pi$ from $D^0$ decay and the charge of the slow $\pi_s$ from $D^{*\pm}$ decay. The RS decay is the sum of the Cabbibo-favored (CF) decay $D^0\to K^-\pi^+$ and $\ddbar$ mixing followed by doubly Cabbibo-suppressed (DCS) decay $\bar{D}^0\to  K^-\pi^+$ where the latter process amplitude can be neglected comparing to the former process. While the WS is sum of two comparable amplitude decay for DCS decay $D^0\to K^+\pi^-$ and $\ddbar$ mixing followed by CF decay $\bar{D}^0\to K^+\pi^-$. Assuming CP conservation and the mixing parameters are small, the time-dependent ratio of WS to RS decay rate is
\begin{eqnarray}
  r_{WS}(t)  =  \Big{(} R_{D} + \sqrt{R_{D}}y^{\prime}\Gamma t+\frac{x^{\prime 2}+y^{\prime 2}}{4}\Gamma^2 t^2\Big{)} e^{-\Gamma t} %\nonumber \\
\end{eqnarray} 
Here the effective mixing parameters $x^{\prime}$ and $y^{\prime}$ are defined in Eqn.(\ref{effxy}) with the strong phase difference $\delta$ between the DCS and CF decay amplitudes.
\begin{eqnarray}
x^{\prime} = x\cos\delta+y\sin\delta, \quad\qquad y^{\prime} = y\cos\delta-x\sin\delta. \label{effxy}%\nonumber
\end{eqnarray} 
\begin{center}
\includegraphics[width=4.2cm]{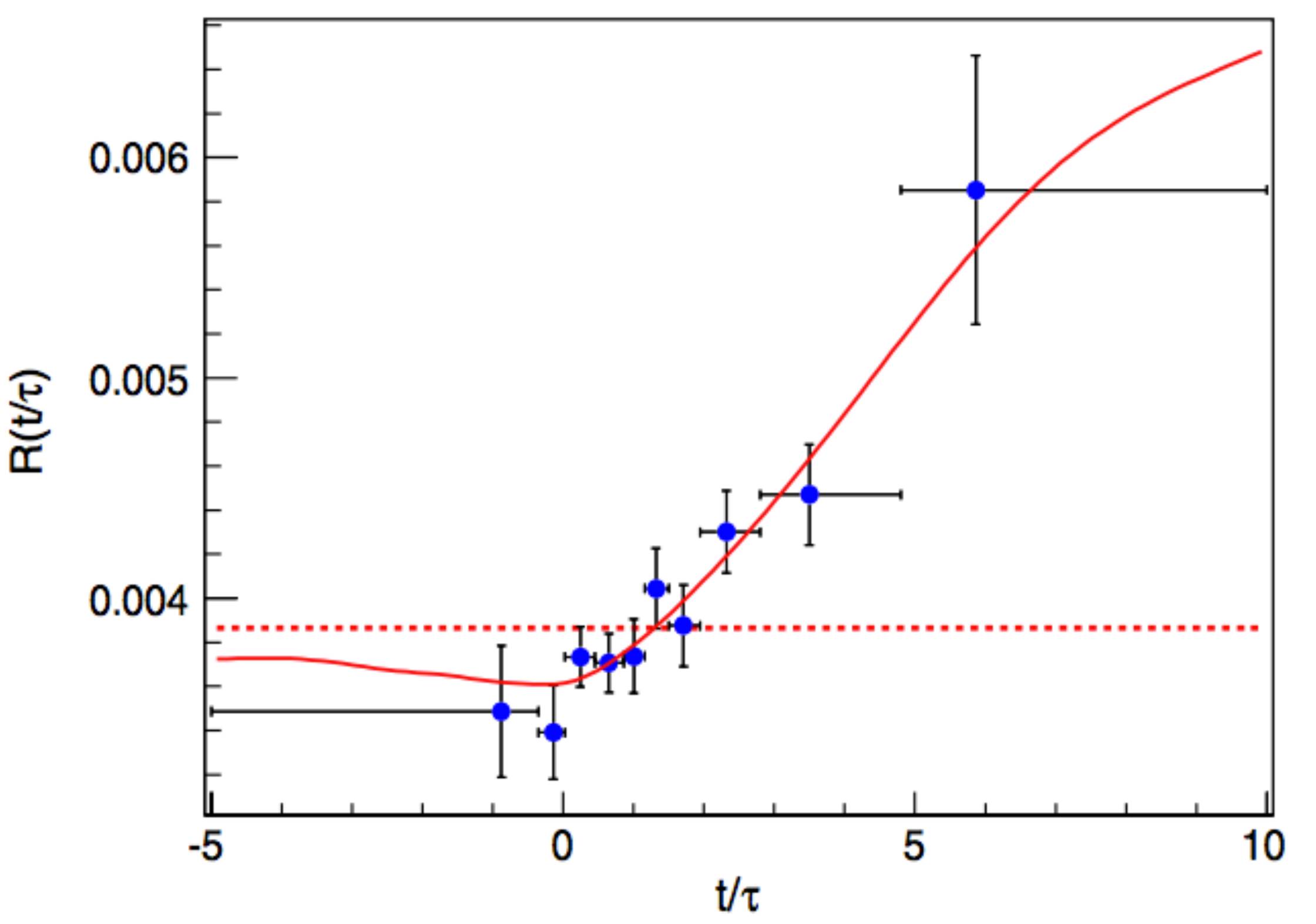}
\includegraphics[width=4cm]{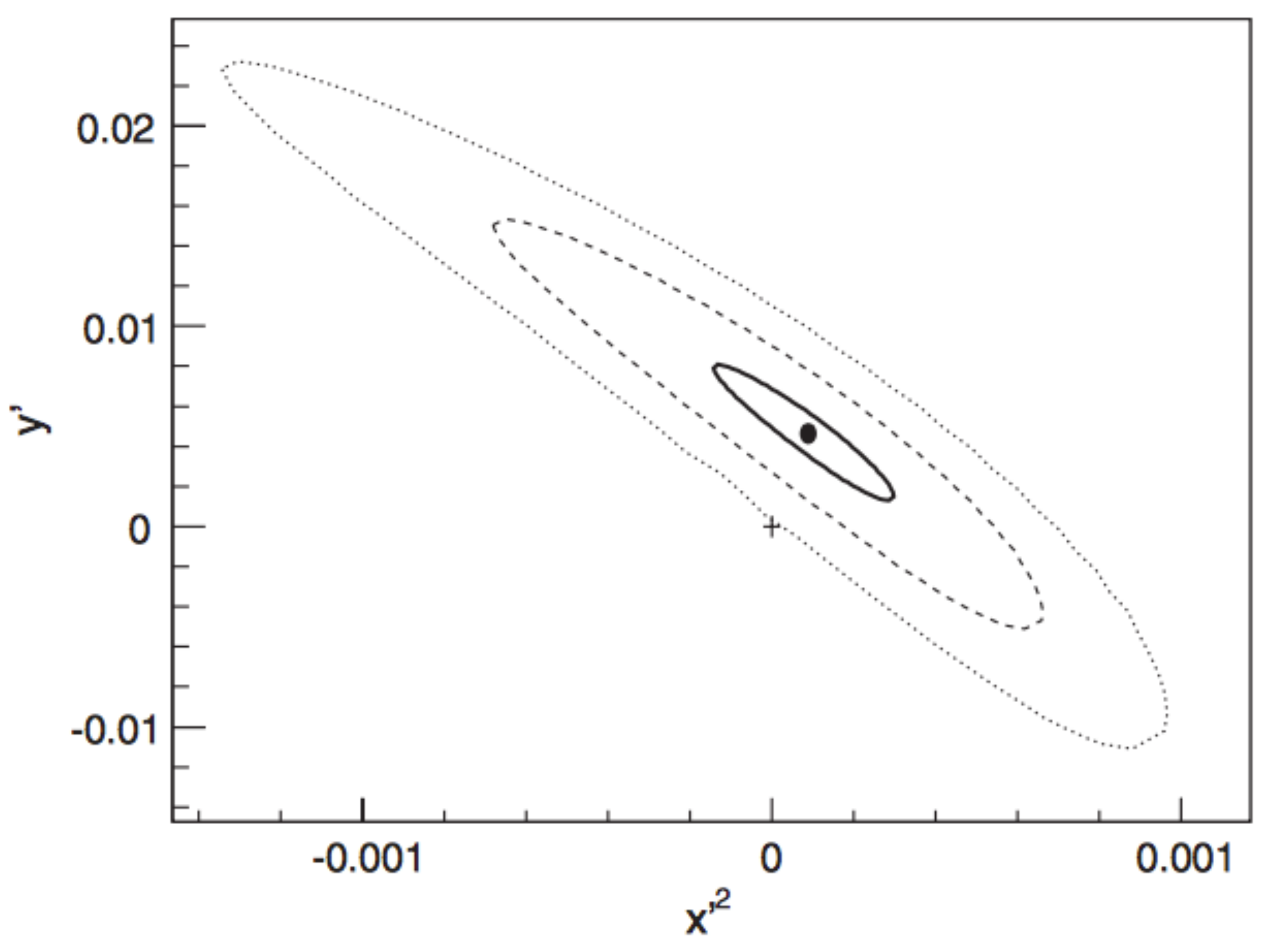}
\figcaption{\label{fig1_2} The left figure show the time-dependent ratio of WS to RS decay ratios fits with(solid) and without(dashed) the mixing hypothesis, the right figure shows the ($x^{\prime2}$, $y^{\prime}$) plane with best-fit point and contour.}
\end{center}

We firstly obtain the ratio of signal and background %, shown in Fig.\ref{fig1_1}, 
by $\Delta M$ fit, where $\Delta M$ is the $D^{*+}-D^0$ mass difference, $\Delta M= M(D^{*+}\to D^0(\to K\pi)\pi_s^+) - M(D^0\to K\pi)$. The measured $D^0$ proper decay time is calcuted as $t=m_{D^0}\vec{L}\cdot\vec{p}/|\vec{p}|^2$ where $\vec{L}$ is the vector joining the decay and production vertices of the $D^0$, $\vec{p}$ is the $D^0$ momentum. 
%\begin{center}
%\includegraphics[width=8cm]{D0Kpi_mq.pdf}
%\figcaption{\label{fig1_1} Time-integrated distribution for the mass difference $\Delta M$ of RS(left) and WS(right).}
%\end{center}

Our fits to the time-dependent ratios of WS to RS decays, after the time resolution thought about, are shown in Fig.\ref{fig1_2}(left) with two hypotheses, with and without mixing. %The results are listed in Table \ref{tab1_1}. 
The $\chi^2$ difference between the "no-mixing" and "mixing" hypotheses, $\Delta\chi^2=\chi^2_{\text{no mixing}}-\chi^2_{\text{mixing}}=33.5-4.2=29.3$ for 2 degrees of freedom, which implies the no-mixing hypothesis is excluded at the 5.1 standard deviation level. We also show $1\sigma$ $3\sigma$ and $5\sigma$ contours around the the best fit point in the ($x^{\prime2}$, $y^{\prime}$) plane shown in Fig.\ref{fig1_2}(right).
%\begin{figure}[!htpb]

%\end{figure}
\begin{comment}
%\begin{table}[!htpb]
\begin{center}
\tabcaption{\label{tab1_1} Time-dependent fitting to r(t) with or without mixing hypothesis, where $dof$ stands for the degrees of freedom. The uncertainties are statistical and systematic combined.}
\footnotesize
\begin{tabular*}{70mm}{@{\extracolsep{\fill}}lcc} 
\toprule \hline %\hline
Test type			& parameters	&  results			\\			
($\chi^2/dof$)		&  			& ($10^{-3}$)		\\ \hline
Mixing (4.2/7)		&  $R_D$		&  $3.53\pm0.13$  	\\
  				&  $y^{\prime}$	&  $4.6\pm3.4$ 	\\
				&  $x^{\prime2}$ & $0.09\pm0.22$  	\\ \cline{2-3}
				& ($x^{\prime2}$, $y^{\prime}$) &  -0.948  \\
Correlation coef. 	& ($R_D$, $x^{\prime2}$) &  +0.737  \\
				& ($R_D$, $y^{\prime}$) &  -0.865  \\  \hline
No mixing (33.5/9)	&  $R_D$		&  $3.864\pm0.059$ \\ \hline 
\bottomrule
\end{tabular*}  
\end{center}
%\end{table}
\end{comment}

\subsection{Self-conjugated decay $D^0\to K_S^0\pi^+\pi^-$\cite{lab2}}
We describle the decay amplitudes for $D^0$ or a $\bar{D}^0$ into the final self-conjugated state $K_S^0\pi^+\pi^-$, $\mathcal{A}_f(\bar{\mathcal{A}}_{f})$, as a function of the Dalitz-plot(DP) variables ($m_+^2$, $m_-^2$)=($m_{K_S^0\pi^+}^2$, $m_{K_S^0\pi^-}^2$). If CP symmetry in the decay is assumed, i.e., $\bar{\mathcal{A}}_f=\mathcal{A}_{\bar{f}}=\mathcal{A}(m_-^2, m_+^2)$, we can derive the time-dependent decay rates for $D^0$ and $\bar{D}^0$ decays to the final state $f$ as 
{\footnotesize{
\begin{eqnarray}
  \begin{split}
    |\mathcal{M}(f,t)|^2 = \Big{\{}(|\mathcal{A}_{f}|^2 + |\frac{q}{p}|^2|\mathcal{A}_{\bar{f}}|^2) \cosh(yt) 
     - 2\Re(\frac{q}{p}\mathcal{A}_{\bar{f}}\mathcal{A}_{f}^{*})\sin(yt) \\
     + (|\mathcal{A}_{f}|^2 - |\frac{q}{p}|^2|\mathcal{A}_{\bar{f}}|^2) \cos(xt)  + 2\Im(\frac{q}{p}A_{\bar{f}}A_{f}^{*})\sin{(xt)}\Big{\}} e^{-t}.
  \end{split}  \\
  \begin{split}
    |\bar{\mathcal{M}}(f,t)|^2 = \Big{\{}(|\mathcal{A}_{\bar{f}}|^2 + |\frac{p}{q}|^2|\mathcal{A}_{f}|^2) \cosh(yt) 
     - 2\Re(\frac{p}{q}\mathcal{A}_{f}\mathcal{A}_{\bar{f}}^{*})\sin(yt) \\
     + (|\mathcal{A}_{\bar{f}}|^2 - |\frac{p}{q}|^2|\mathcal{A}_{f}|^2) \cos(xt)  + 2\Im(\frac{p}{q}A_{f}A_{\bar{f}}^{*})\sin{(xt)}\Big{\}} e^{-t}.
  \end{split}   
\end{eqnarray}
}}here the unit of time $t$ is $D^0$ lifetime $\tau$. if no CPV allowed, $|q/p|=1$ and $\arg(q/p)=0$.

We analyse a data sample of 921 $fb^{-1}$ recorded at or near $\Upsilon(4S)$ and at $\Upsilon(5S)$ resonances. We reconstruct the $D^0$ mesons through the decay chain $D^{*+}\to D^0(\to K_S^0\pi^+\pi^-)\pi_s^+$, where the charge of $\pi_s$ is used to tag the flavor of the $D$ meson. We determine the signal yield from a two-dimensional fit to $M-Q$ distribution, where $M=M_{K_S^0\pi^+\pi^-}$ is the $D^0$ invariant mass and $Q=M_{K_S^0\pi^+\pi^-\pi_s}-M_{K_S^0\pi^+\pi^-}-m_{\pi_s}$ is the kinetic energy released in $D^*$ decay.
%\begin{center}
%\includegraphics[width=8cm]{D0ks2p_sigbg.pdf}
%\figcaption{\label{fig1}   Figure 1. }
%\end{center}

For DP model, we use 12 resonances described by relativistic Breit-Wigner parameterizations with mass-dependent widths, and Zemach tensors for angular dependence for the P- and D-wave decays. For $\pi\pi$ S-wave dynamics, we adopt the K-matrix formalism with P-vector approximation. For $K_S^0\pi$ S-wave, we use LASS model at production experiment with an effective range non-resonant component. 

In time-dependent Dalitz plot fitting, we set the free parameters to be ($x$, $y$), the $D^0$ lifetime $\tau$, the time resolution function parameters and the amplitude model parameters. We extract the mixing parameters $x=(0.56\pm0.19)\%$ and $y=(0.30\pm0.15)\%$ with the statistical correlation coefficient between $x$ and $y$ of 0.12. The DP distribution and its Dalitz variables projections are shown in Fig.\ref{fig2} and the $D^0$ proper time projection is shown in Fig.\ref{fig2_2}. 

We also search for CPV with CPV parameters $|q/p|$ and $\arg(q/p)$ included in the fit. The values for the mixing parameters from the this fit are essentially identical to the ones from the CP-conserved fit. The resulting CPV parameters are $|q/p|=0.90^{+0.16}_{-0.15}$ and $\arg(q/p)=(-6\pm11)^{o}$. We consider serval contributions to the experimental systematic uncertainty. By exploring the negative log-likelihood distribution on the plane of mixing parameters, we draw the two-dimensional ($x$, $y$) confidence level contours for both the CP-conserved and CPV-allowed fits shown in Fig.\ref{fig2_2}(right). The final fit results for CP-conserved and CPV-allowed both are listed in Table.\ref{tab2}.
\begin{center}
\includegraphics[width=8cm]{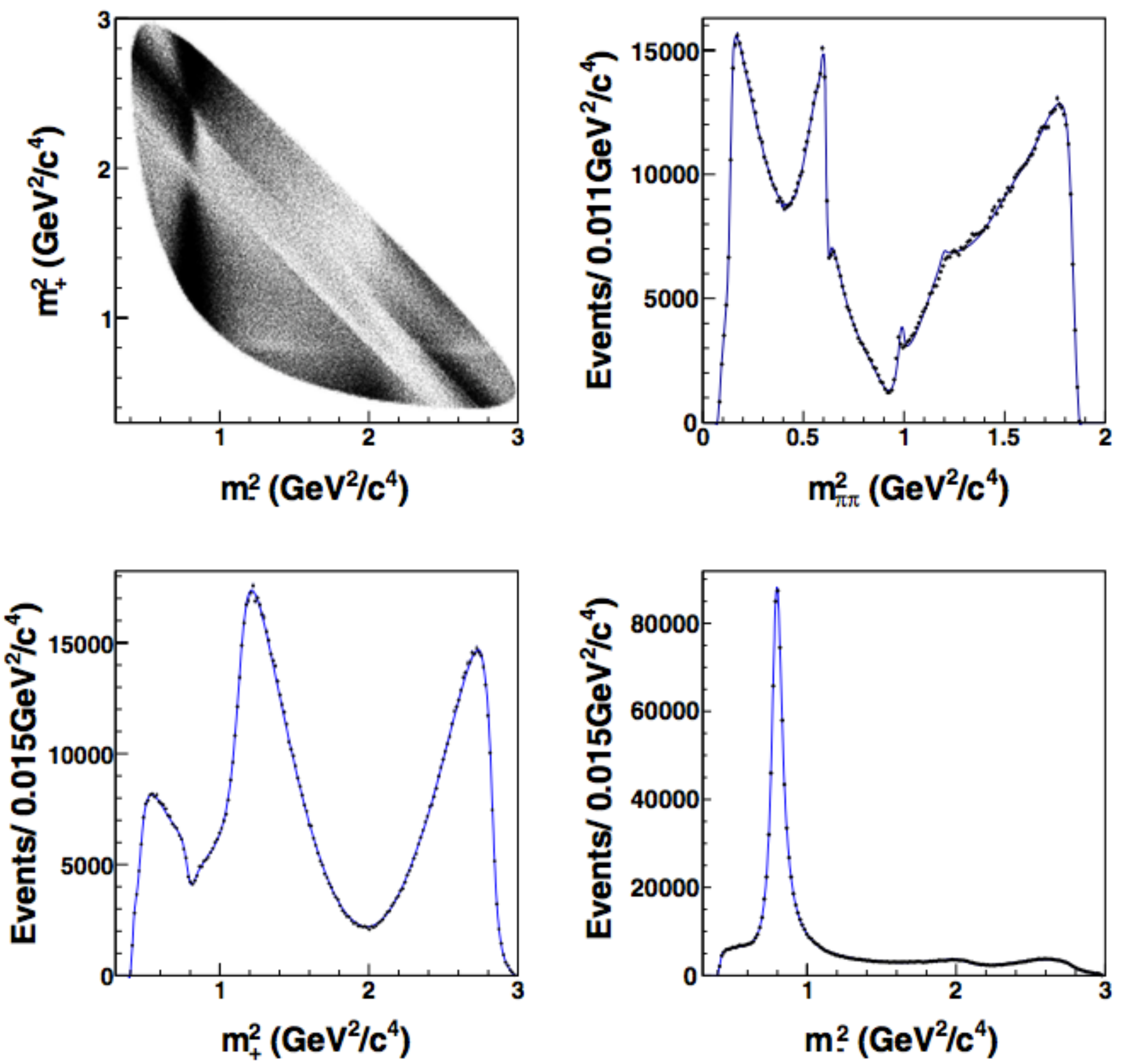} 
\figcaption{\label{fig2}Dalitz distribution and Dalitz variables projections for the selected data sample.}
\end{center}

\begin{center}
\includegraphics[width=3.8cm]{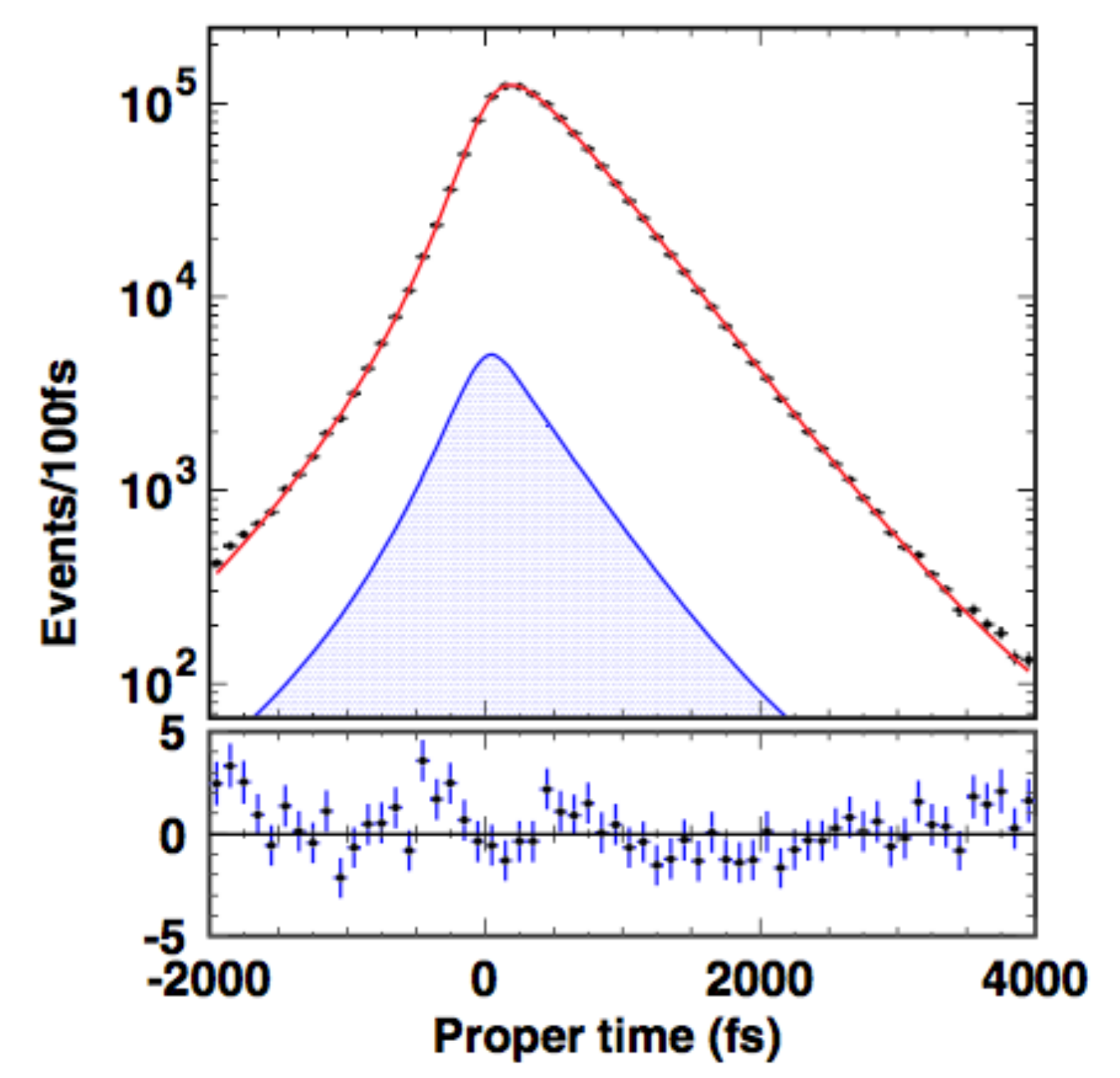}
\includegraphics[width=4.2cm]{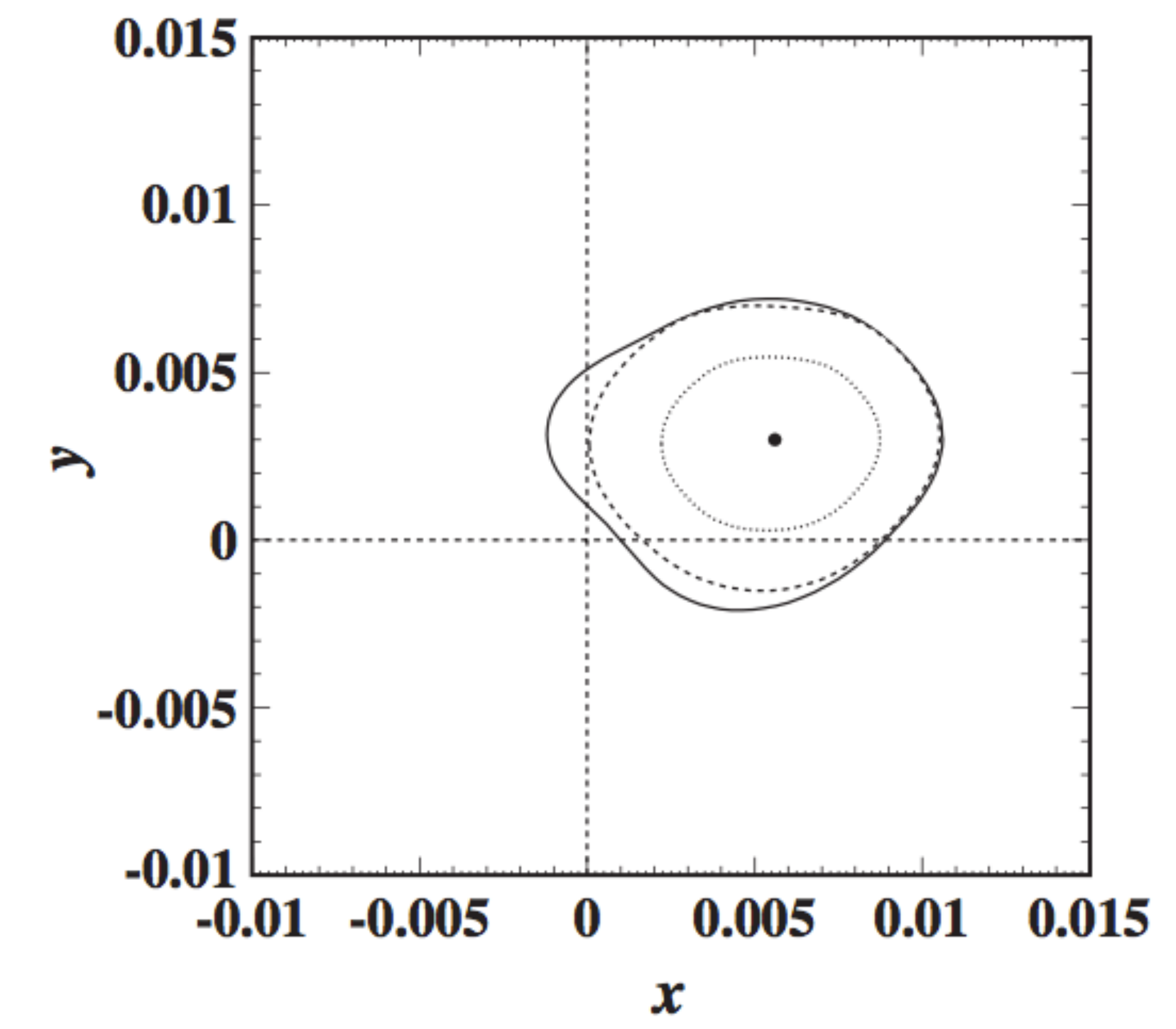}
\figcaption{\label{fig2_2}The left figure is the proper time distribution in signal region for CP-conserved and the right for the confidence level contours of mixing parameters for CP-conserved Dalitz fit(dashed) and CP-allowed fit(solid).}
\end{center}
%\begin{comment}
\begin{center}
\tabcaption{\label{tab2} Fit results for the mixing parameters $x$ and $y$ from the CP-conserved fit and the CPV-allowed fit.}
\footnotesize
\begin{tabular*}{70mm}{@{\extracolsep{\fill}}lcc}
\toprule \hline %\hline
Fit type	&  Parameter	&  Fit result	\\ \hline
\multirow{2}{*}{No CPV}	& x(\%)	&  $0.56\pm0.19^{+0.03+0.06}_{-0.09-0.09}$ \\
					& y(\%)	&  $0.30\pm0.15^{+0.04+0.03}_{-0.05-0.06}$ \\ \hline
\multirow{4}{*}{CPV}		& x(\%)	&   $0.56\pm0.19^{+0.03+0.06}_{-0.09-0.09}$ \\					
					& y(\%)	&  $0.30\pm0.15^{+0.04+0.03}_{-0.05-0.06}$ \\ 
					& $|q/p|$	&  $0.90^{+0.16+0.05+0.06}_{-0.15-0.04-0.05}$ \\	
					& $\arg(q/p)$ & $-6\pm11\pm3^{+3}_{-4}$  \\ 	
\hline %\hline
\bottomrule
\end{tabular*}  
\end{center} 
%\end{comment}

\subsection{CP eigenstate states $D^0\to K^+K^-/\pi^+\pi^-$\cite{lab3}}
Belle measured of $D^0-\bar{D}^0$ mixing in decays to CP eigenstates $K^+K^-/\pi^+\pi^-$ based on the total Belle data sample of 976 $fb^{-1}$. Mixing in $D^0$ decays to CP eigenstates, such as $D^0\to K^+K^-$, manifests in a lifetime that differs from the lifetime of decays to flavor eigenstates, such as $D^0\to K^-\pi^+$. The quantity $y_{CP}=\frac{\tau(K^-\pi^+)}{\tau(K^+K^-)} -1$ is equal to the mixing parameters $y$ if CP is conserved. If the  CP is violated, the lifetime of $D^0$ and $\bar{D}^0$ decaying to the same CP eigenstates also differ and the lifetime asymmetry, defined as $A_{\Gamma}=\frac{\tau(\bar{D}^0\to K^-K^+) - \tau(D^0\to K^+K^-)}{\bar{D}^0\to K^-K^+) + \tau(D^0\to K^+K^-)}$ becomes non-zero.

\begin{center}
\includegraphics[width=4cm]{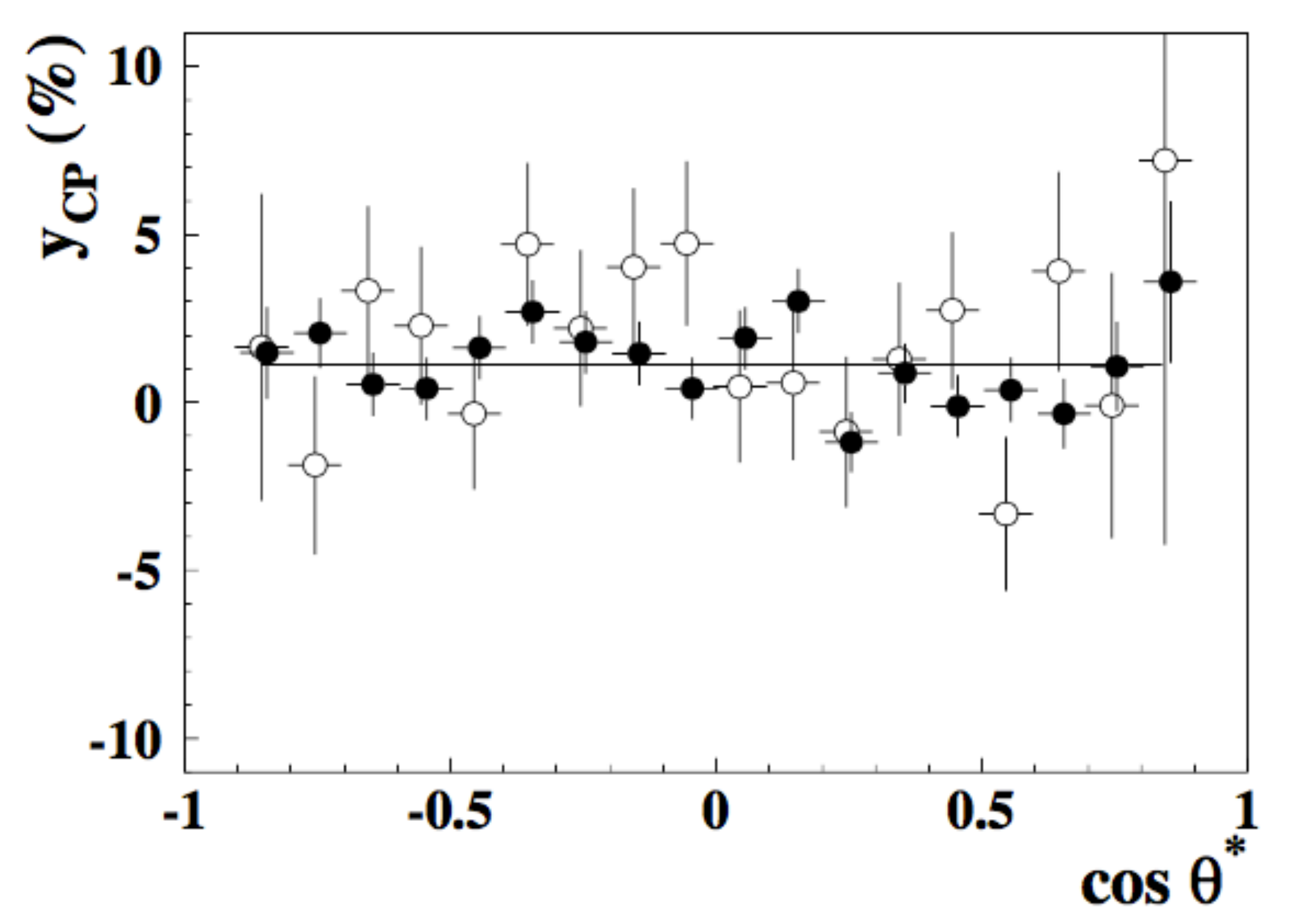}
\includegraphics[width=4cm]{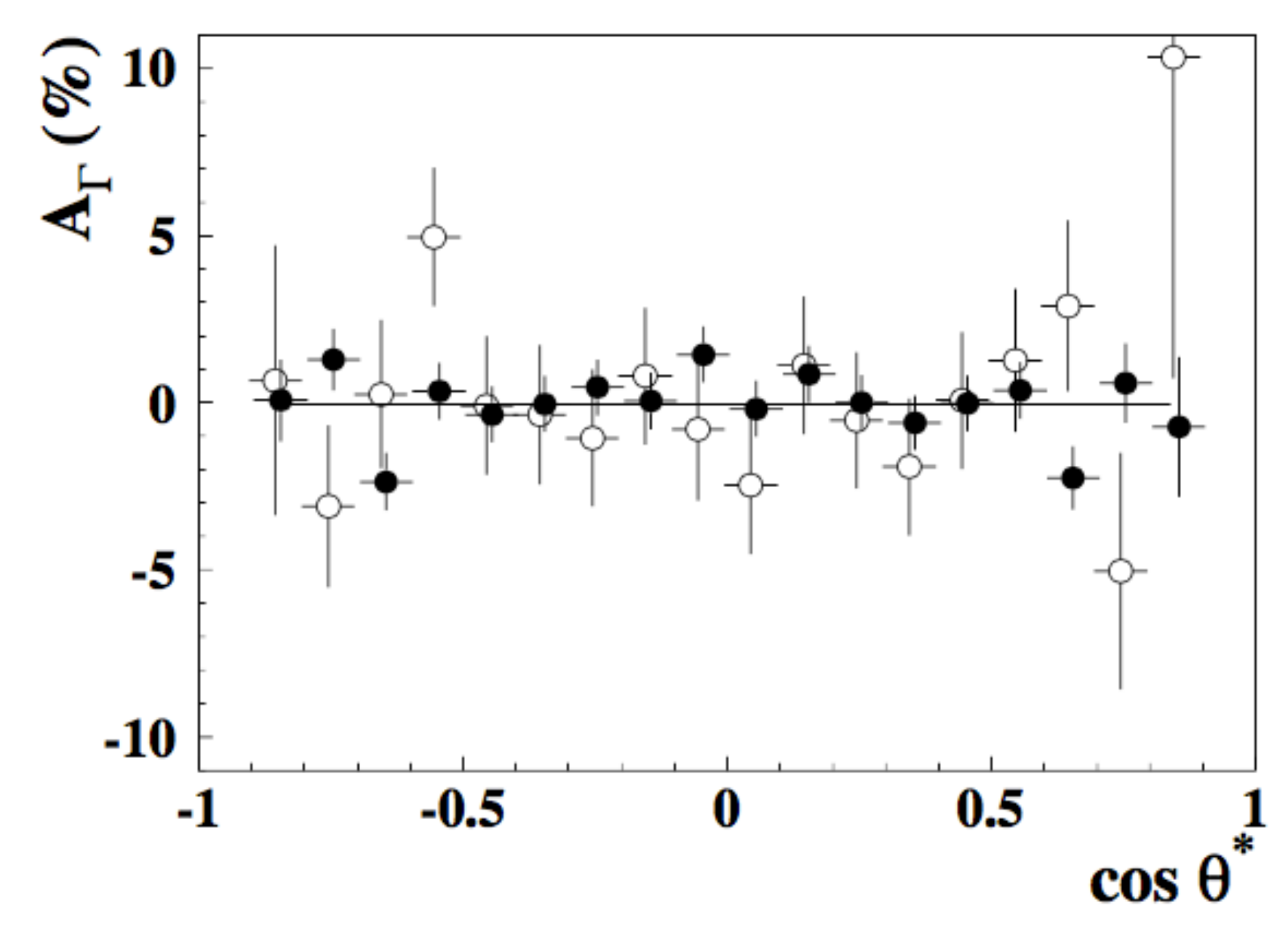}
\figcaption{\label{fig3}Fitted $y_{CP}$ and $A_{\Gamma}$ in bins of $\cos\theta^*$ for 3-layer SVD data(open circles) and for 4-layer SVD data(full circles). The horizontal line is the result of fitting the points to a constant. }
\end{center}

To extract $y_{CP}$ and $A_{\Gamma}$, the decay modes are fitted simultaneously in each $\cos\theta^*$ bin and separately for each of the two SVD configurations. The fitted values of $y_{CP}$ and $A_{\Gamma}$ in bins of $\cos\theta^*$ are shown in Fig.\ref{fig3}. The values obtained with a least-squares fit to a constant are  
$y_{CP}=(+1.11\pm0.22\pm0.09)\%$ with a significance of $4.7\sigma$ and $A_{\Gamma}=(-0.03\pm0.20\pm0.07)\%$.

\subsection{Search for CPV in $D^0\to\pi^0\pi^0$ decay\cite{lab4}}
In the SM, CPV in singly Cabbibo-suppressed (SCS) charm decays arises due to interference between the tree and loop(penguin) amplitudes and is suppressed to $\sim10^{-3}$. In this section, we report the measurement of the time-integrated CP-violating asymmetry($A_{CP}$) in $D^0\to\pi^0\pi^0$ and the update of $D^0\to K_S^0\pi^0$ with an integrated luminosity of 960 $fb^{-1}$. The measured asymmetry
\begin{eqnarray}
A_{rec} = \frac{N_{rec}^{D^{*+}\to D^0\pi^+_{s}} - N_{rec}^{D^{*-}\to \bar{D^0}\pi^-_{s}}}{ N_{rec}^{D^{*+}\to D^0\pi^+_{s}} + N_{rec}^{D^{*-}\to \bar{D^0}\pi^-_{s}}}  %\nonumber  
\end{eqnarray}
where $N_{rec}$ is the number of reconstructed signal events, includes three contributions: the underlying CP asymmetry $A_{CP}$, the forward-backward asymmetry($A_{FB}$) due to $\gamma-Z^0$ interference in $e^+e^-\to c\bar{c}$ and higher-order QED effects, and the detection asymmetry between positively and negatively charged pion($A_{\epsilon}^{\pi_s}$). The last contribution depends on the transverse momentum $p_T^{\pi_s}$ and polar angle $\theta^{\pi_s}$ of the slow pion and is independent of the final state. We thus extract $A_{CP}$ and $A_{FB}$ using
\begin{eqnarray}
  A_{CP}=\frac{1}{2}\Big{(}A_{rec}^{cor}(\cos\theta^{*})+A_{rec}^{cor}(-\cos\theta^{*})\Big{)} \\%\nonumber \\
  A_{FB}=\frac{1}{2}\Big{(}A_{rec}^{cor}(\cos\theta^{*})-A_{rec}^{cor}(-\cos\theta^{*})\Big{)} %\nonumber
\end{eqnarray}
and from the weighted average over the $|\cos\theta^*|$ bins, we obtained asymmetry shown in Fig.\ref{fig4}.
\begin{center}
\includegraphics[width=8cm]{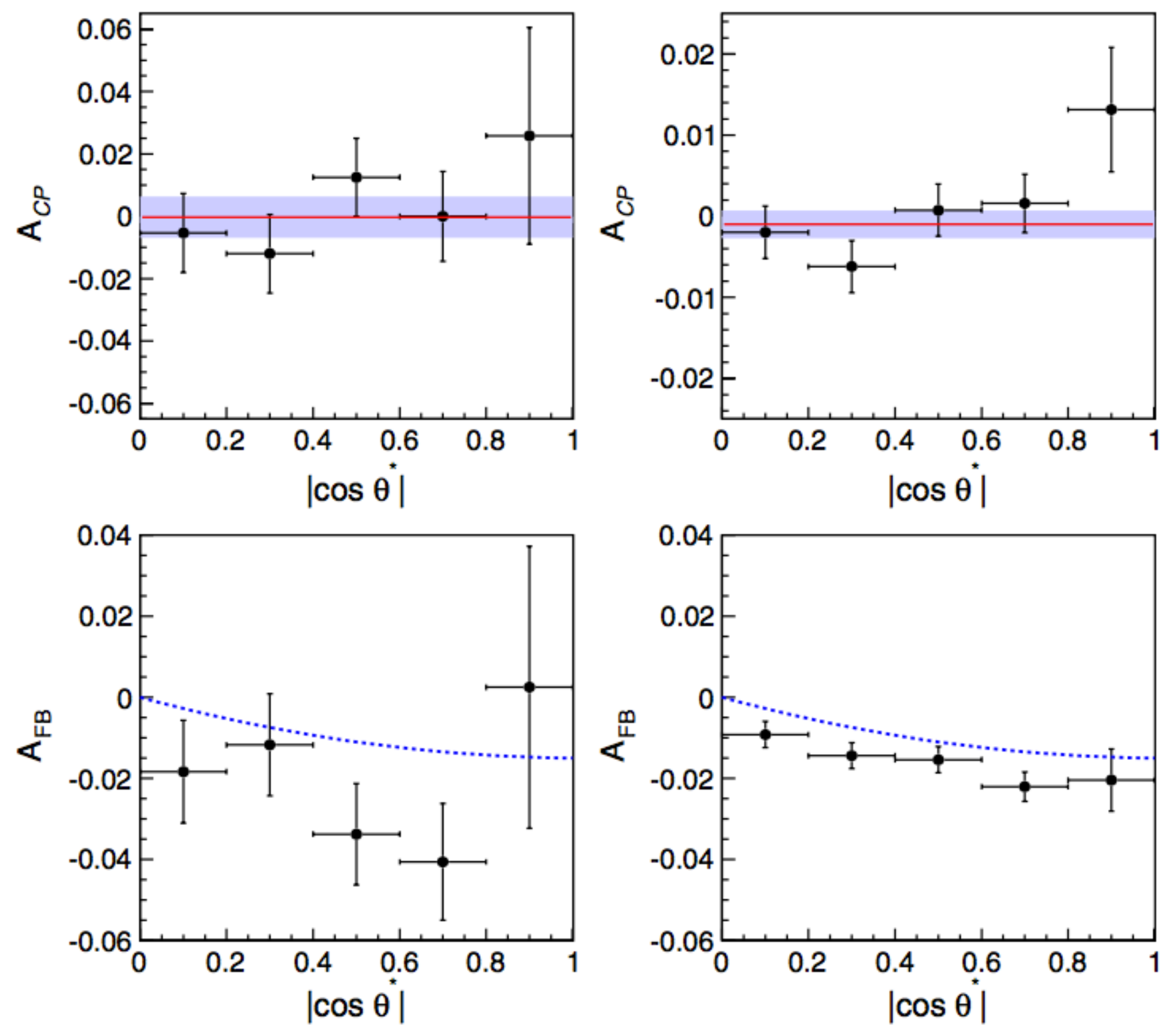}
\figcaption{\label{fig4} CP violation asymmetry $A_{CP}$ (top) and forward-backward asymmetry $A_{FB}$ (bottom) values as a function of $|\cos\theta^*|$. }
\end{center}

We identify three significant sources of systematic uncertainty, for detail see Ref. \cite{lab8}. Thus the asymmetry obtained in the rate of $D^0$ and $\bar{D}^0$ decays to $\pi^0\pi^0$ final state: $A_{CP}(D^0\to\pi^0\pi^0)=(-0.03\pm0.64\pm0.10)\% $, is consistent with no CP violation. This result constitutes an order of magnitude improvement over the exisiting result. We also present an updated measurement of the CP asymmetry in $D^0\to K_S^0\pi^0$ decay: $ A_{CP}(D^0\to K_{S}^0\pi^0)=(-0.21\pm0.16\pm0.07)\% $.

\begin{comment}
\begin{center}
\tabcaption{\label{tab4} Summary of systematic uncertainties ($\%$) in $A_{CP}$.}
\footnotesize
\begin{tabular*}{70mm}{@{\extracolsep{\fill}}lcc}
\toprule \hline %\hline
  Source			&	$\pi^0\pi^0$	&	$K_{S}^{0}\pi^0$	\\ \hline
  Signal shape		& 	$\pm0.03$	&	$\pm0.01$	\\
  Slow pion correction	&	$\pm0.07$	& 	$\pm0.07$	\\
  $A_{CP}$ extraction method 	& $\pm0.07$	& $\pm0.02$	\\
  $K^0$/$\bar{K^0}$-material effects & 	&	$\pm0.01$	\\ \hline
  Total				&	$\pm0.10$	&  	$\pm0.07$	\\
\hline %\hline
\bottomrule
\end{tabular*}  
\end{center}  
\end{comment}
 
\section{Search for $D^0$ rare decay $D^0\to\gamma\gamma$\cite{lab5}}
Flavor changing neutral current (FCNC) processes are forbidden at tree level in SM although they can occur at higher orders. The rare decay $D^0\to\gamma\gamma$, mediated by a $c\to u\gamma\gamma$ transition, has very small short distance contributions but there can be large long-distance effects owing to the contributions of intermediate vector mesons. We search for $D^0\to\gamma\gamma$ using an 832.4 $fb^{-1}$ of data sample collected near the $\Upsilon(4S)$ and $\Upsilon(5S)$ resonances.

To reduce large combinatorial backgrounds arising from random photon combinations, we require that the $D^0$ be produced in the decay $D^{*+}\to D^0\pi^+$. The $D^0\to\gamma\gamma$ branching fraction is thus measured with respect to  a well measured mode $D^0\to K_S^0\pi^0$ using the following relation:
\begin{eqnarray}
\mathcal{B}(D^0\to\gamma\gamma)=\frac{(N/\epsilon)_{D^0\to\gamma\gamma}}{(N/\epsilon)_{D^0\to K_S^0\pi^0}}\times\mathcal{B}(D^0\to K_S^0\pi^0)_{WA}.
\end{eqnarray}
where $N$ and $\epsilon$ are the signal yield and detection efficiency of the respective channels.

Using the two-dimensional fit of $M(\gamma\gamma)$ and $\Delta M$, we find $4\pm15$ signal, $210\pm32$ peaking and $2934\pm59$ combinatorial background events, respectively, shown in Fig.\ref{fig5}. In absence of a statistically significant signal, we derive an upper limit at $90\%$ CL on the signal yield $N_{UL}^{90\%}=25$ accounting systematic uncertainty and its corresponding branching fraction at $8.4\times10^{-7}$ which is the most restrictive upper limit on $D^0\to\gamma\gamma$ to data and is approaching the SM prediction. This FCNC decay will be probed further at the next-generation flavor factories such as $\BTWO$.
\begin{center}
\includegraphics[width=8cm]{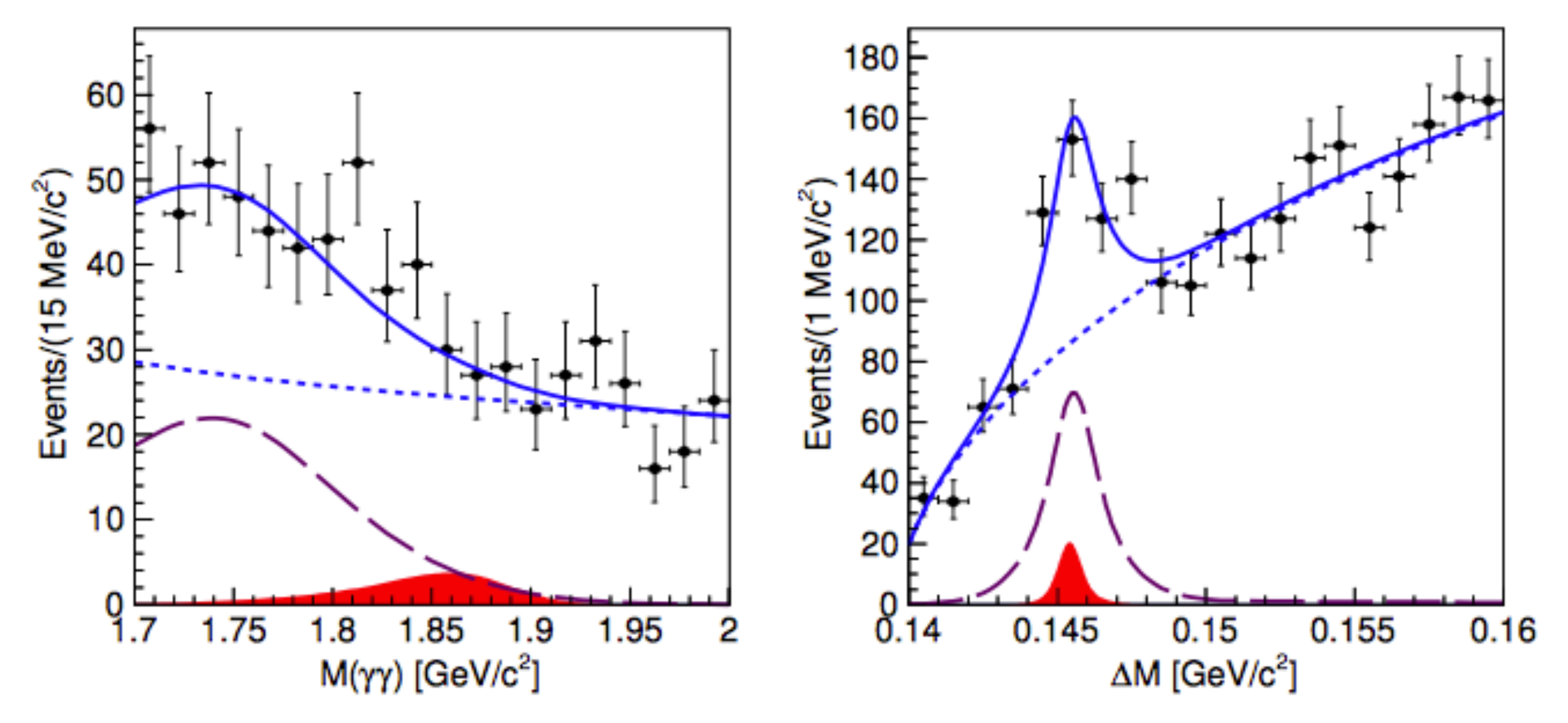}
\figcaption{\label{fig5}Projections of candidate events onto $M(\gamma\gamma)$(left) and $\Delta M$(right) applying a signal-region criterion on the other variable. Points with error bars are the data, blue solid curves are the results of fit, blue dotted curves represent the combinatorial background. Red filled histograms show the signal where its yield is scaled up to corresponding $90\%$ CL upper limit value.}
\end{center}

\begin{comment}
\begin{center}
\tabcaption{\label{tab2} estimation of upper limit on B.R.}
\footnotesize
\begin{tabular*}{70mm}{@{\extracolsep{\fill}}lc}
\toprule \hline %\hline
  Quantity						&  Value	\\ \hline
  $N_{UL}^{90\%}(D^0\to\gamma\gamma)$	&  24  	\\
  $N(D^0\to K_S^0\pi^0)$			&  343050  	\\
  $\epsilon_{K_S^0\pi^0}$			&  $7.2\%$	\\
  $\epsilon_{\gamma\gamma}$		&  $7.3\%$	\\
  BR($K_{S}^0\pi^0$)			&  $1.19\times10^{-2}$  \\ \hline
  BR($D^0\to\gamma\gamma$)		&  $<8.1\times10^{-7}$  \\
\hline %\hline
\bottomrule
\end{tabular*}  
\end{center}
\end{comment}

\begin{comment}
\end{multicols}
\begin{center}
\tabcaption{\label{tab2} Status of charmed baryon spectroscopy in PDG2014\cite{PDG2014}. }
\footnotesize
\begin{tabular*}{170mm}{@{\extracolsep{\fill}}c|cccc|cc}
\toprule \hline %\hline
Exp./ Theo.	& Belle 	& BABAR  	& $\BES$ 	&  CLEO  & SM	  & MSSM 	\\ \hline
Data($fb^{-1}$)	& 833 	& 470.5		&  2.92	& 13.8 	&  \multirow{2}{6.5em}{mainly via long-distance} 	  &	\multirow{2}{4.5em}{exchange of gluinos}	\\
Publish Year	& preliminary &  \href{http://journals.aps.org/prd/abstract/10.1103/PhysRevD.85.091107}{2012}	& 2015	& \href{http://journals.aps.org/prl/abstract/10.1103/PhysRevLett.90.101801}{2003}	&        &		\\
B.R.($\times10^{-6}$) & 	$<0.81$	& $<2.2$	& $<3.8$	& $<29$	& $1\%\sim3\%$	& $\sim6$	\\ 
\hline %\hline
\bottomrule
\end{tabular*}  
\end{center}
\begin{multicols}{2}
\end{comment}

\section{Charmed baryons spectroscopy}
Charmed baryons studies have of great interest physics, like the di-quark correlation in light quarks from single charmed baryon; the $QQ$ potential, similar to $Q\bar{Q}$ in charmonium, from doubly charmed baryons.
%Charmed baryons spectroscopy referred to PDG\cite{PDG2014} are shown in Table \ref{tab2}. 
%Most of them 
Most of the charmed baryons listed in PDG\cite{PDG2014} 
are observed by CLEO except that $\Sigma_c(2800)$ $\Xi_c(2980)$ and $\Xi_c(3080)$ observed by Belle;  $\Lambda_c(2940)^+$ $\Xi_c(3055)$ and $\Omega_c(2770)$ observed by BaBar. Their spin-parity are almost from quark-model prediction. $\Xi_c(3055)$ has not yet been included into PDG, while confirmed by Belle\cite{lab7}. 

\begin{comment}
\begin{center}
\tabcaption{\label{tab2} Status of charmed baryon spectroscopy in PDG2014\cite{PDG2014}. }
\footnotesize
\begin{tabular*}{85mm}{@{\extracolsep{\fill}}ll|ll|ll} 
\toprule \hline %\hline
 $cdu$	   			&  $I(J^P)$  		&  $cdu$	   			& $I(J^P)$   		&  $csu$			&  $I(J^P)$  				\\ \hline
 $\Lambda_c^+$ 		& $0(\frac{1}{2}^+)$  & $\Sigma_c(2455)$ 		& $1(\frac{1}{2}^+)$  & $\Xi_c$ 			& $\frac{1}{2}(\frac{1}{2}^+)$  	\\
 $\Lambda_c(2595)^+$ 	& $0(\frac{1}{2}^+)$  & $\Sigma_c(2520)$ 		& $1(\frac{3}{2}^+)$  & $\Xi_c^{\prime}$	& $\frac{1}{2}(\frac{1}{2}^+)$  	\\
$\Lambda_c(2625)^+$ 	& $0(\frac{1}{2}^+)$  & $\Sigma_c(2800)$ 		& $1({?}^{?})$  		& $\Xi_c(2645)$ 	& $\frac{1}{2}(\frac{3}{2}^+)$  	\\
$\Lambda_c(2880)^+$ 	& $0(\frac{1}{2}^+)$  &					&   				&  $\Xi_c(2790)$	& $\frac{1}{2}(\frac{1}{2}^-)$  	\\
$\Lambda_c(2940)^+$ 	& $0({?}^{?})$  	     	&  					&   				&  $\Xi_c(2815)$	& $\frac{1}{2}(\frac{3}{2}^-)$  	\\ \cline{3-4}
  					&  				& $css$				& $I(J^P)$			& $\Xi_c(2980)$	& $\frac{1}{2}({?}^{?})$  		\\ \cline{3-4}
  					&  				& $\Omega_c$			& $0(\frac{1}{2}^+)$ 	& $\Xi_c(3055)$	& $\frac{1}{2}({?}^{?})$  		\\       
  					&  				& $\Omega_c(2770)$	& $0(\frac{3}{2}^+)$	& $\Xi_c(3080)$	& $\frac{1}{2}({?}^{?})$  		\\ 
\hline %\hline
\bottomrule
\end{tabular*}  
\end{center}
\end{comment}

\subsection{Charmed strange baryons studies\cite{lab6}}
We report a search for doubly charmed baryon $\Xi_{cc}^{+(+)}$ with the $\Lambda_c^+ K^-\pi^+(\pi^+)$ in Fig.\ref{fig6_1} and $\Xi_c^0\pi^+(\pi^+)$ in Fig.\ref{fig6_2}(\ref{fig6_3}) final states using a 980 $fb^{-1}$ data sample. No significant signal of $\Xi_{cc}$ is observed in all invariant mass distributions.

\begin{center}
\includegraphics[width=6cm]{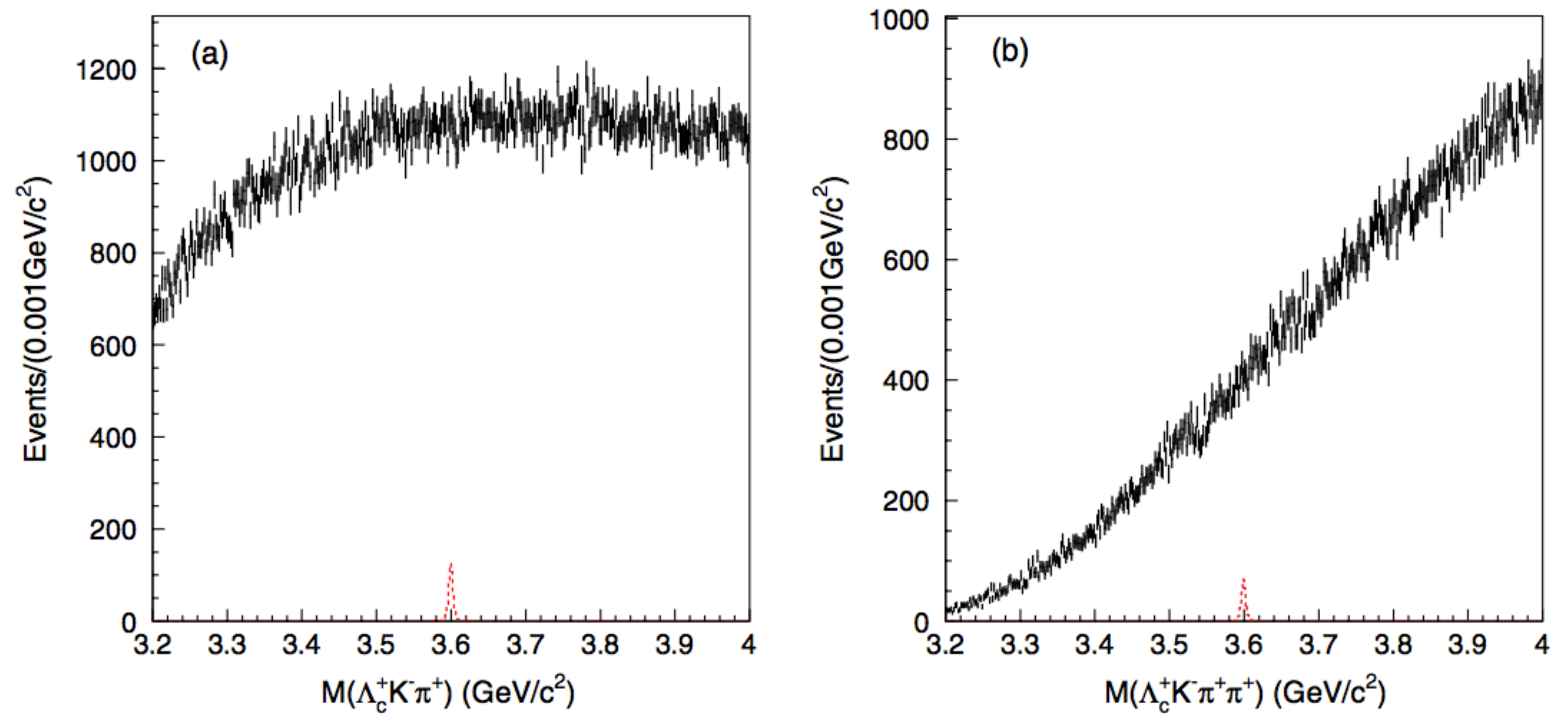}
\figcaption{\label{fig6_1} Invariant mass distribution of $\Xi_{cc}$ candidates for (a) $M(\Lambda_c^+K^-\pi^+)$ and (b) $M(\Lambda_c^+K^-\pi^+\pi^+)$.}
\end{center}
\begin{center}
\includegraphics[width=5.4cm]{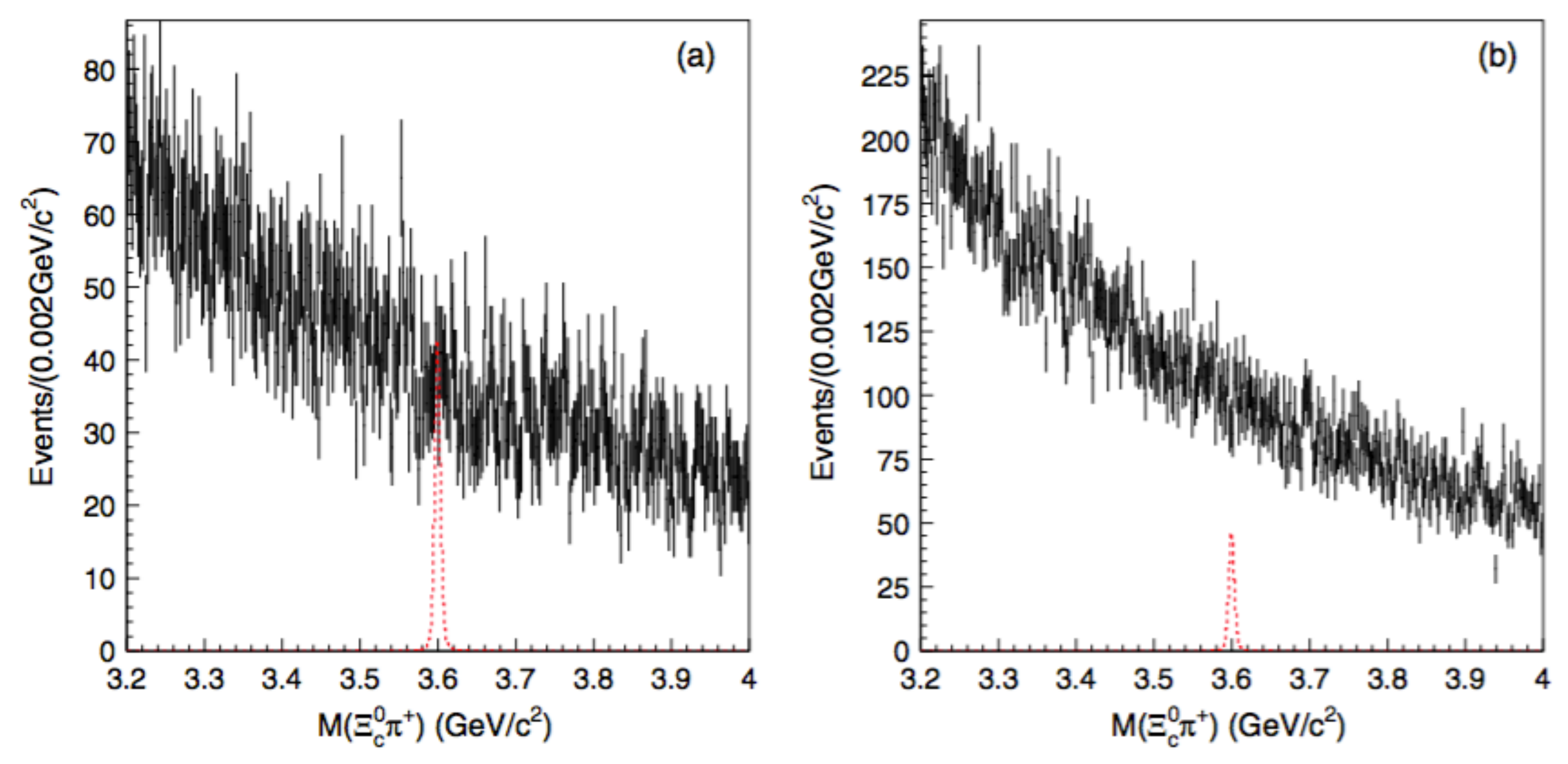}
\includegraphics[width=2.7cm]{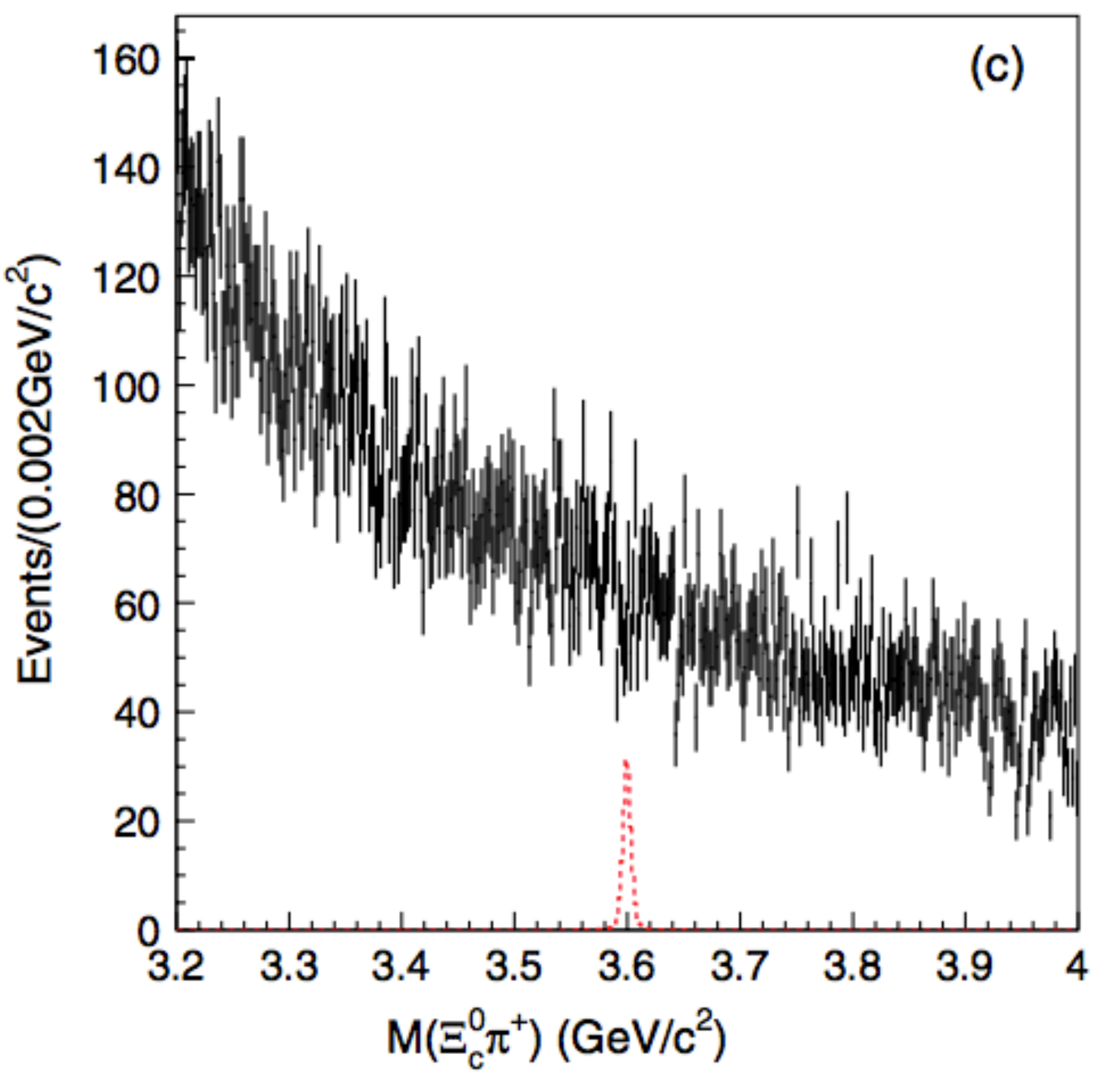}
\figcaption{\label{fig6_2} $M(\Xi_c^0\pi^+)$ distribution in the $\Xi_{cc}^+$ search region for $\Xi_c^0\to$ (a) $\Xi^-\pi^+$, (b) $\Lambda K^-\pi^+$, (c) $pK^-K^-\pi^+$.}
\end{center}
\begin{center}
\includegraphics[width=5.4cm]{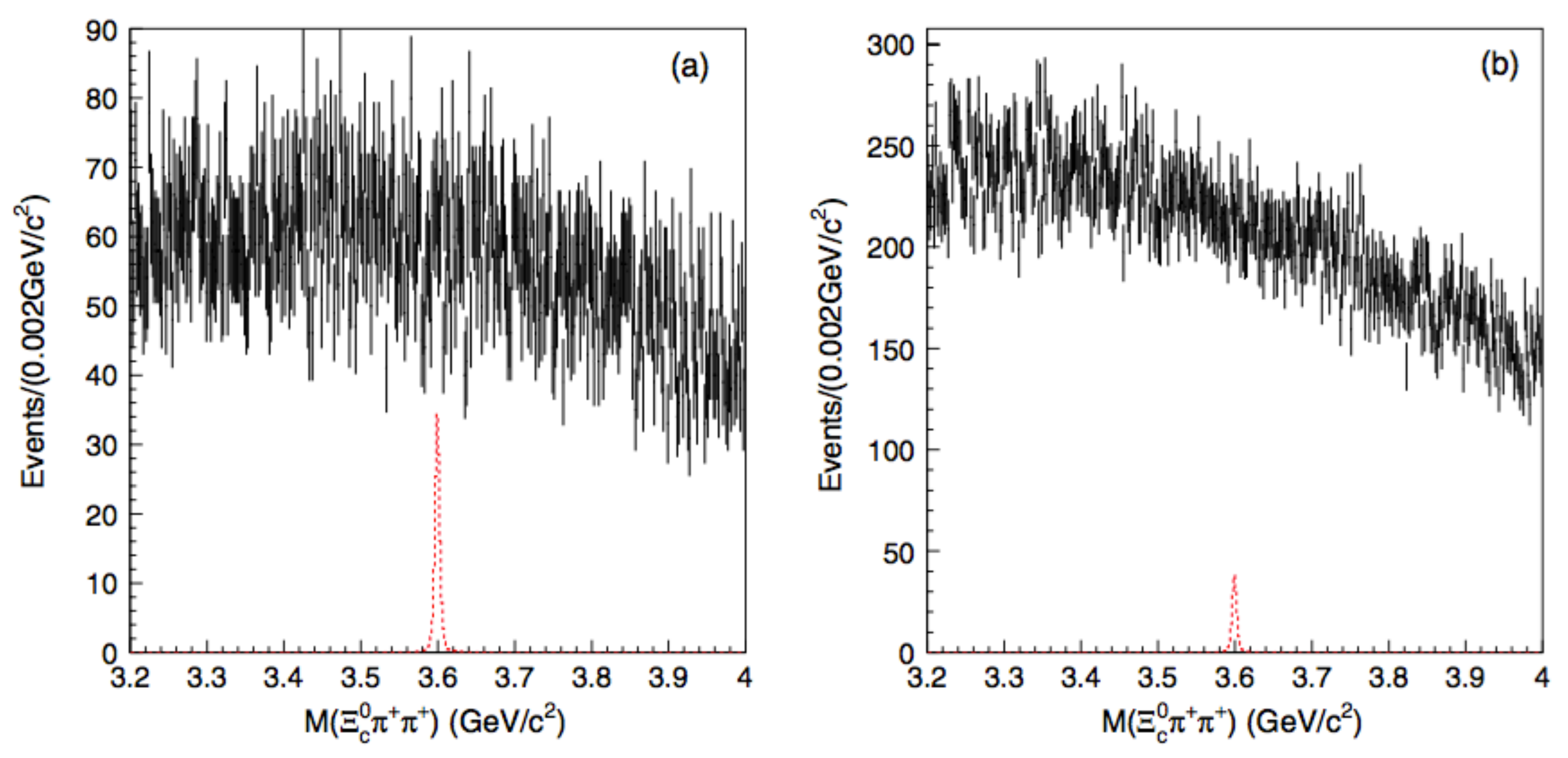}
\includegraphics[width=2.7cm]{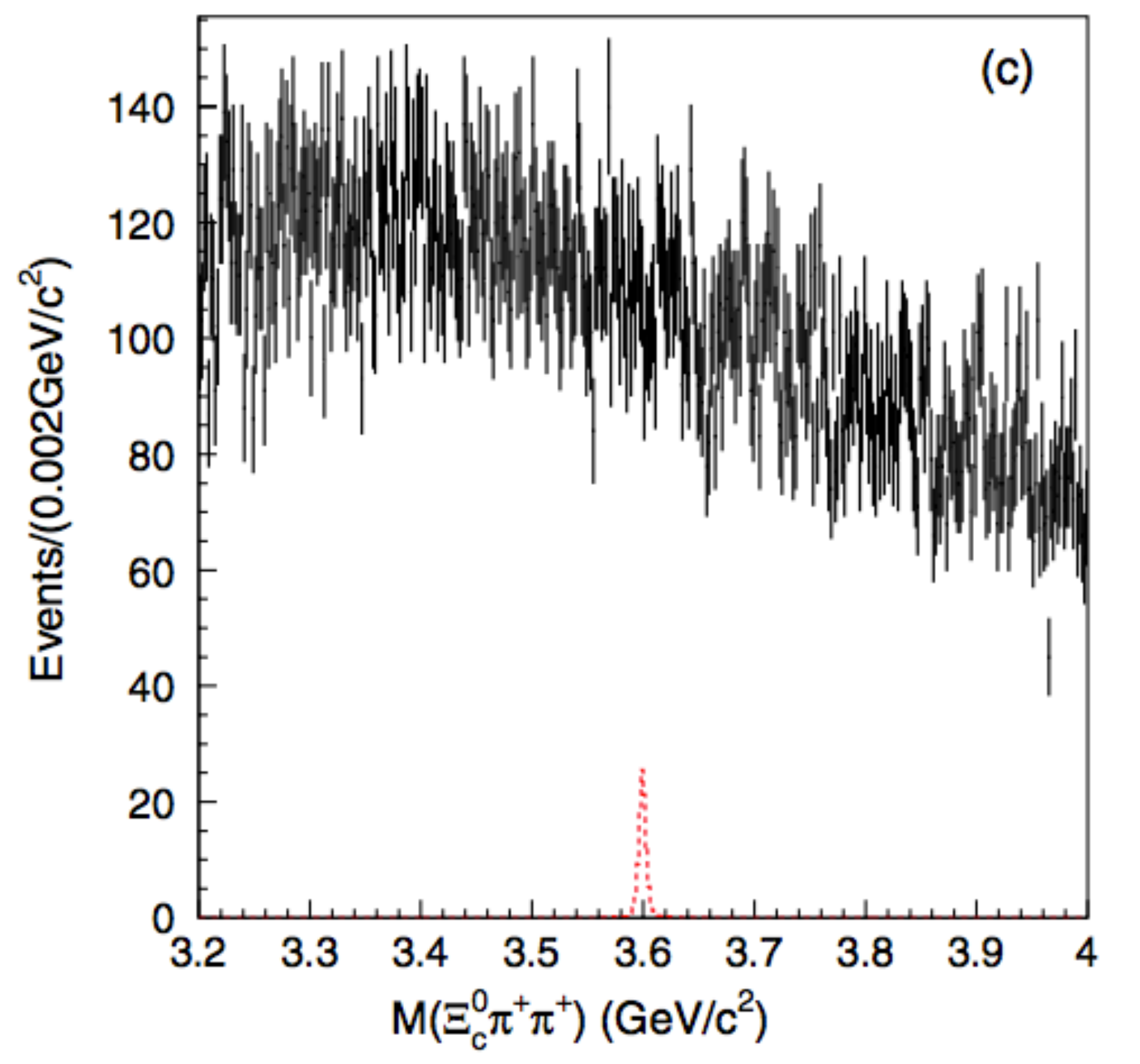}
\figcaption{\label{fig6_3} $M(\Xi_c^0\pi^+\pi^+)$ distribution in the $\Xi_{cc}^{++}$ search region for $\Xi_c^0\to$ (a) $\Xi^-\pi^+$, (b) $\Lambda K^-\pi^+$, (c) $pK^-K^-\pi^+$.}
\end{center}

We also search for two exited charmed strange baryons $\Xi_c(3055)^+$ and $\Xi_c(3123)^+$ with the $\Sigma_c^{++}(2455)K^-$ and $\Sigma_c^{++}(2520)K^-$ final states, see Fig.\ref{fig6_4}. The $\Xi_c(3055)^+$ signal is observed with a significance of 6.6 standard deviations including systematic uncertainty, while no signature of the $\Xi_c(3123)^+$ is seen.
\begin{center}
\includegraphics[width=8cm]{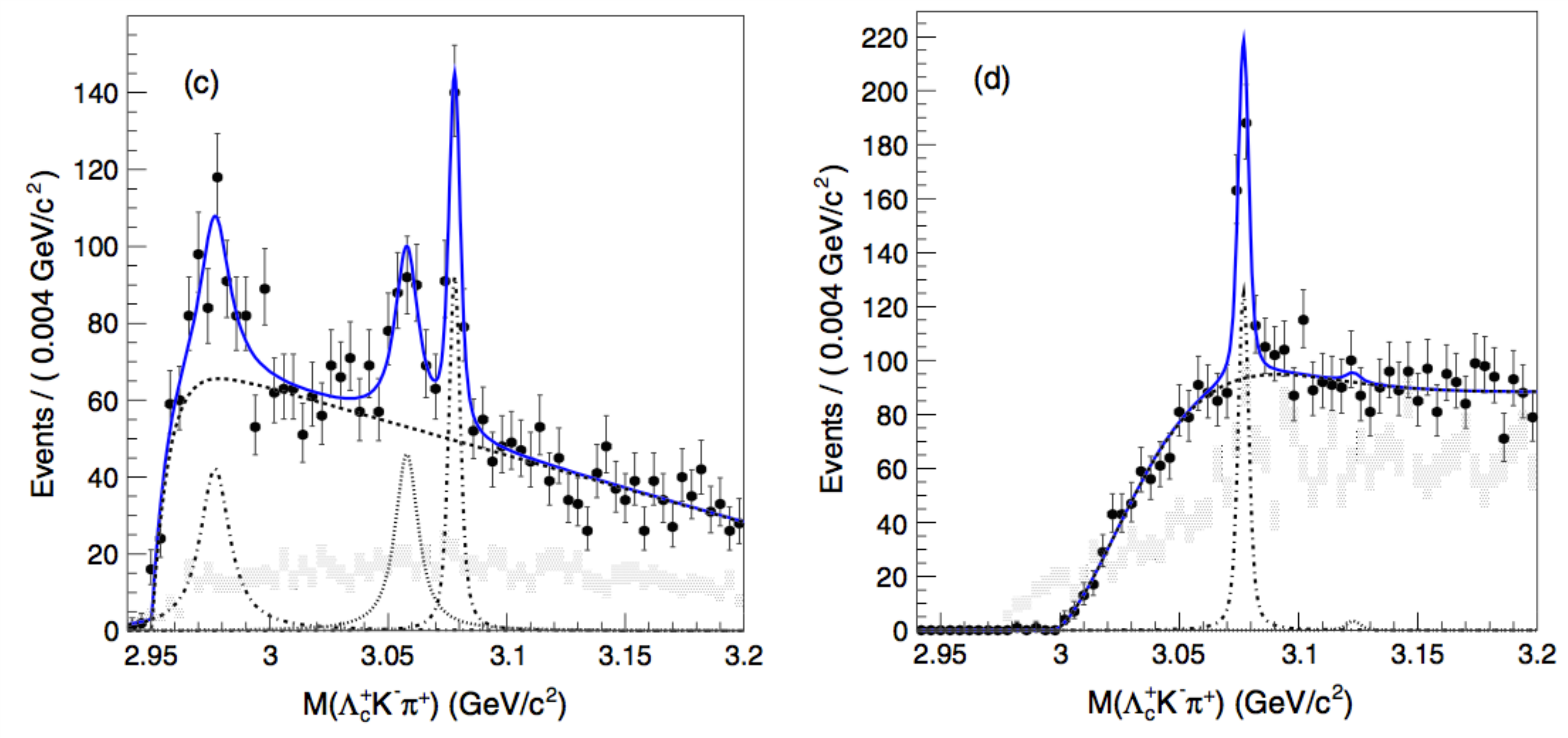}
\figcaption{\label{fig6_4} The $M(\Lambda_c^+K^-\pi^+)$ distribution with $\Sigma_c(2455)^{++}$ selection (left) and the $M(\Lambda_c^+K^-\pi^+)$ distribution with $\Sigma_c(2520)^{++}$ selection (right). }
\end{center}

We also study properties of the $\Xi_c(2645)^+$ and measure a width of $2.6\pm0.2\pm0.4$ MeV/$c^2$, see Fig.\ref{fig6_5}, which is the first significant determination.
\begin{center}
\includegraphics[width=9cm]{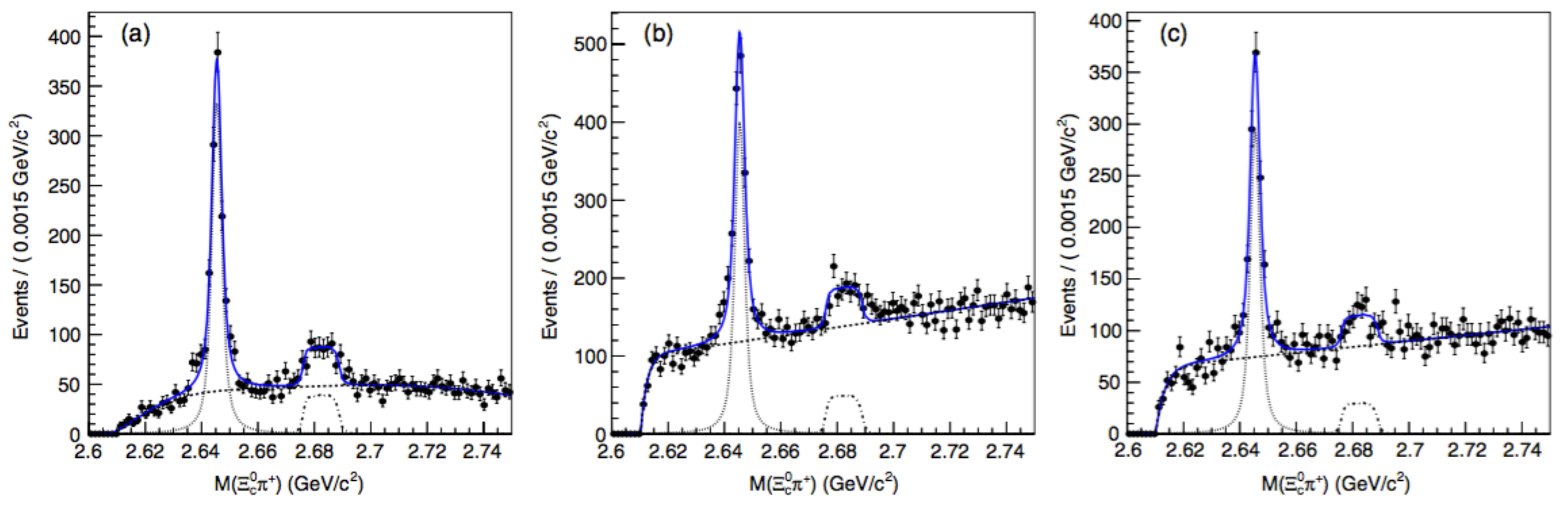}
\figcaption{\label{fig6_5} $M(\Xi_c^0\pi^+)$ distribution below the $\Xi_{cc}^+$ search region for $\Xi_c^0\to$ (a) $\Xi^-\pi^+$, (b) $\Lambda K^-\pi^+$, (c) $pK^-K^-\pi^+$.}
\end{center}

\subsection{Precise measurement of mass and width of $\Sigma_c(2455)$ and $\Sigma_c(2520)$\cite{lab7}}
We present the measurement of the masses and widths of the baryons states $\Sigma_c(2455)^{0/++}$ and $\Sigma_c(2520)^{0/++}$ using a $\Upsilon(4S)$ data sample of 711 $fb^{-1}$. The result for the mass differences (with respect to the $\Lambda_c^+$ mass) and the decay widths of $\Sigma_c(2455)^{0/++}$ and $\Sigma_c(2520)^{0/++}$, shown in Fig.\ref{fig7}, are summarized in Table \ref{tab7} with factor four improvement for mass measurement.

We also calculate the mass splittings $M_0({\Sigma_c^{++}}-M_0{\Lambda_c^0}$ from $\Delta M_0(\Sigma_c^0)$ and $\Delta M_0(\Sigma_c^{++})$ as $0.22\pm0.01\pm0.01$ MeV/$c^2$ for $\Sigma_c(2455)$ and $0.01\pm0.15\pm0.03$ MeV/$c^2$ for $\Sigma_c(2520)$. These measurements are the most precise to data and confirm the mass split with more than $10\sigma$.
\begin{center}
\includegraphics[width=8cm]{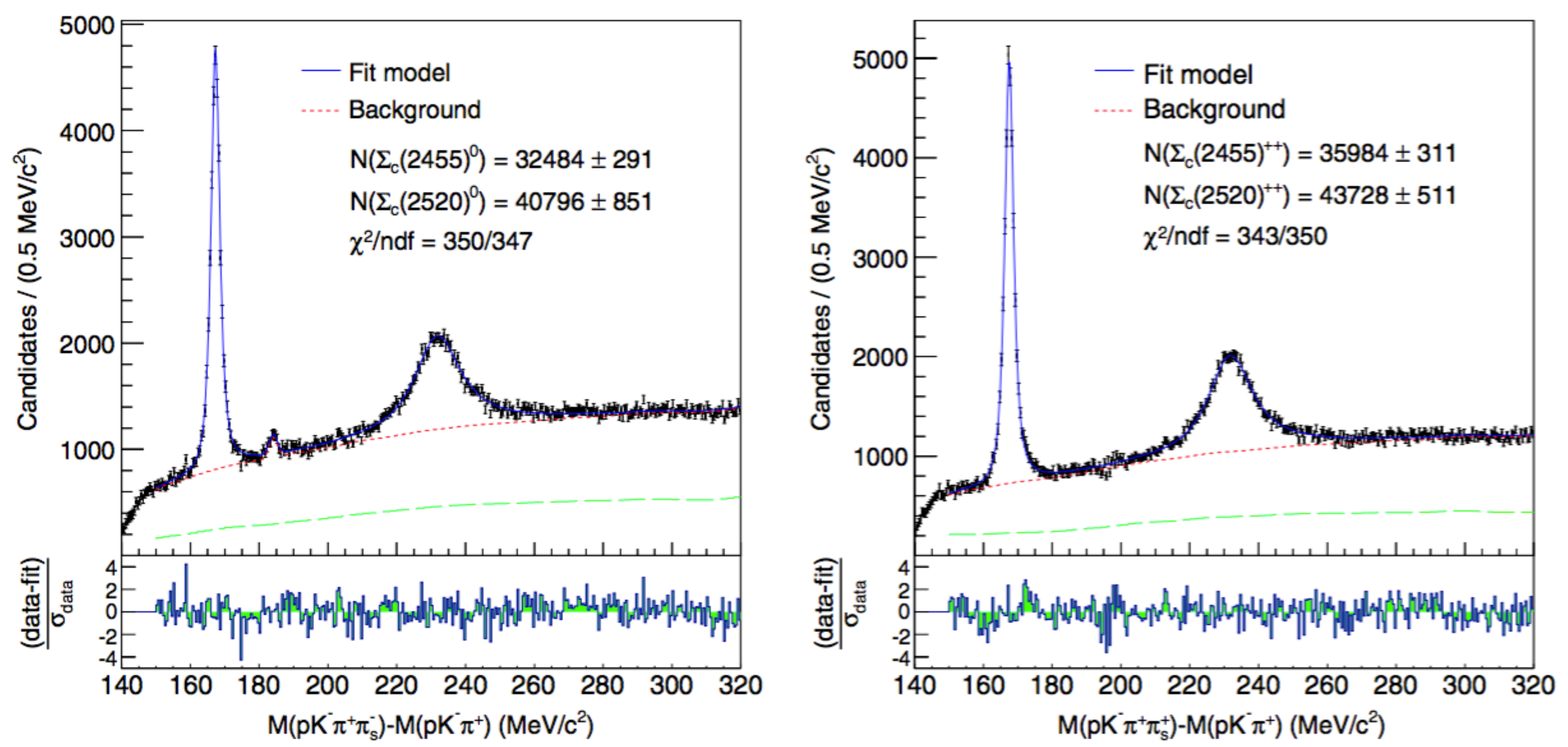}
\figcaption{\label{fig7} Fits to the mass differences $M(pK^-\pi^+\pi_s^{-/+})-M(pK^-\pi^+)$ (left/right) obtained from data.}
\end{center}
\begin{center}
\tabcaption{\label{tab7} The measurements of the masses and the widths of the $\Sigma_c(2455)^{0/++}$ and $\Sigma_c(2520)^{0/++}$. }
\footnotesize
\begin{tabular*}{80mm}{@{\extracolsep{\fill}}l|cc}
\toprule \hline %\hline
			& $M_{\Sigma_c}-M_{\Lambda_c}$ ($MeV/c^2$)		&  $\Gamma(MeV/c^2)$   \\ \hline
$\Sigma_c(2455)^0$	 	& $167.29\pm0.01\pm0.02$	&  $1.76\pm0.04^{+0.09}_{0.21}$	\\
$\Sigma_c(2455)^{++}$	& $167.51\pm0.01\pm0.02$	&  $1.84\pm0.04^{+0.07}_{0.21}$	\\
$\Sigma_c(2455)^0$	 	& $231.98\pm0.11\pm0.04$	&  $15.41\pm0.41^{+0.20}_{0.32}$    \\
$\Sigma_c(2520)^{++}$ 	& $231.99\pm0.10\pm0.02$	&  $14.77\pm0.25^{+0.18}_{0.30}$	\\ 
\hline %\hline
\bottomrule
\end{tabular*}  
\end{center}

\subsection{Absolute branch ratio of $\Lambda_c^+\to pK^-\pi^+$\cite{lab8}}
We present the first model-independent measurement of the absolute branching fraction of the $\Lambda_c^+\to pK^-\pi^+$ decay using a data sample of 978 $fb^{-1}$. The number of $\Lambda_c^+$ baryons is determined by reconstructing the recoiling $D^{(*)-}\bar{p}\pi^+$ system in events of the type $e^+e^-\to D^{(*)-}\bar{p}\pi^+\Lambda_c^+$, see Fig.\ref{fig8}.

The absolute branching fraction of $\Lambda_c^+\to pK^-\pi^+$ decay is given by 
\begin{eqnarray}
\mathcal{BR}(\Lambda_c^+\to pK^-\pi^+)=\dfrac{N(\Lambda_c^+\to pK^-\pi^+)}{N_{inc}^{\Lambda_c}f_{bias}\epsilon(\Lambda_c^+\to pK^-\pi^+)} 
\end{eqnarray}
where $N_{inc}^{\Lambda_c}$ in the number of inclusively reconstructed $\Lambda_c^+$ baryons, $f_{bias}$ takes into account potential dependence of the inclusive $\Lambda_c^+$ reconstruction efficiency on the $\Lambda_c^+$ decay mode. 

The branch fraction is measured to be $\mathcal{B}(\Lambda_c^+\to pK^-\pi^+)=(6.84\pm0.24^{+0.21}_{-0.27})\%$, which represents a fivefold improvement in precision over previous model-dependent determinations. This measurement will also improve significantly the precision of the branching fraction of other $\Lambda_c^+$ decays and of decays of $b-$flavored mesons and baryons involving $\Lambda_c^+$.
\begin{center}
\includegraphics[width=6cm]{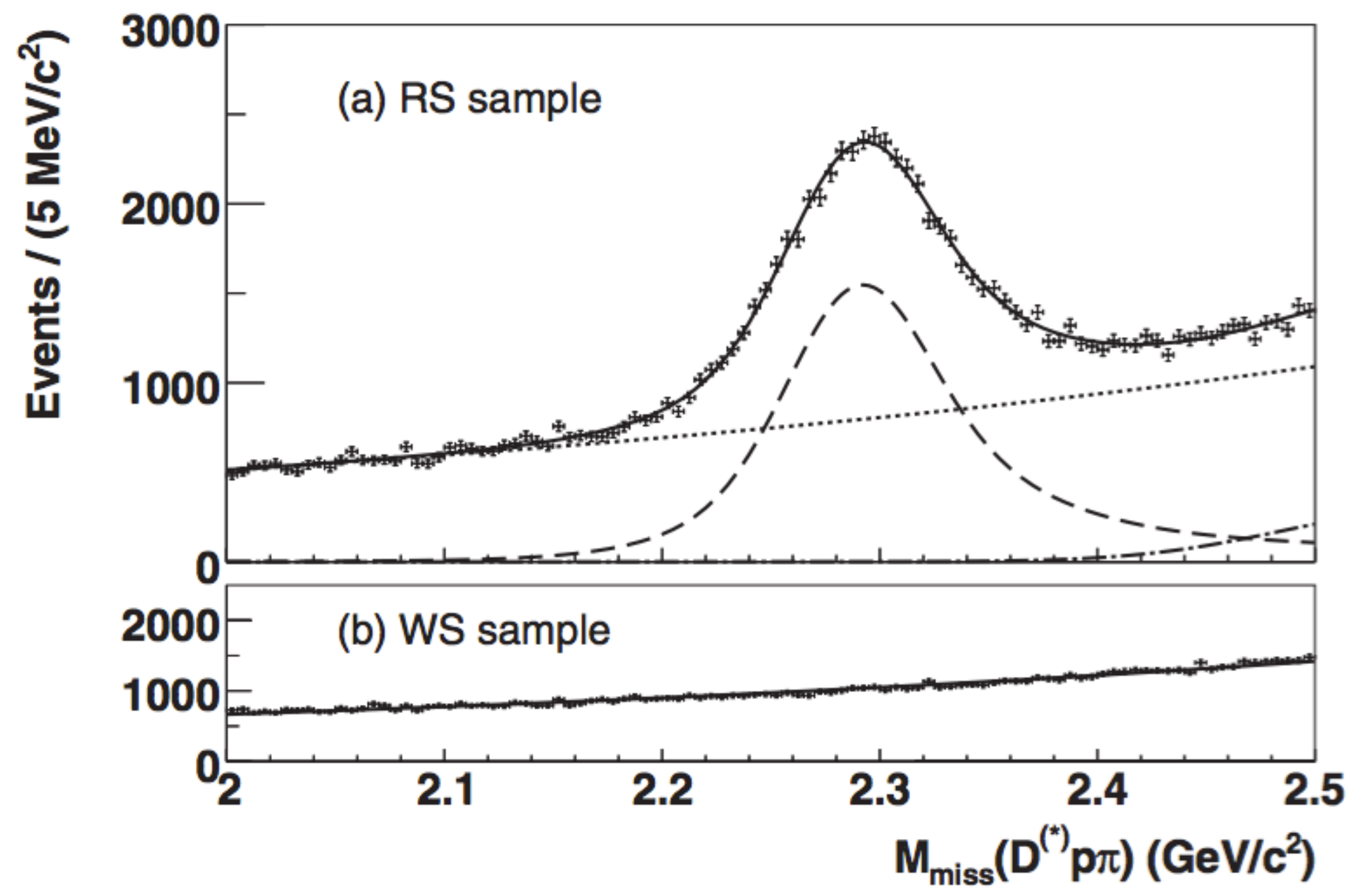}
\figcaption{\label{fig8} The $M_{miss}(D^{(*)}\bar{p}\pi)$ data distribution for inclusively reconstructed $\Lambda_c^+$ baryons from the (a) RS and (b) WS samples with superimposed fit results (slide line).}
\end{center}
\begin{center}
\includegraphics[width=7cm]{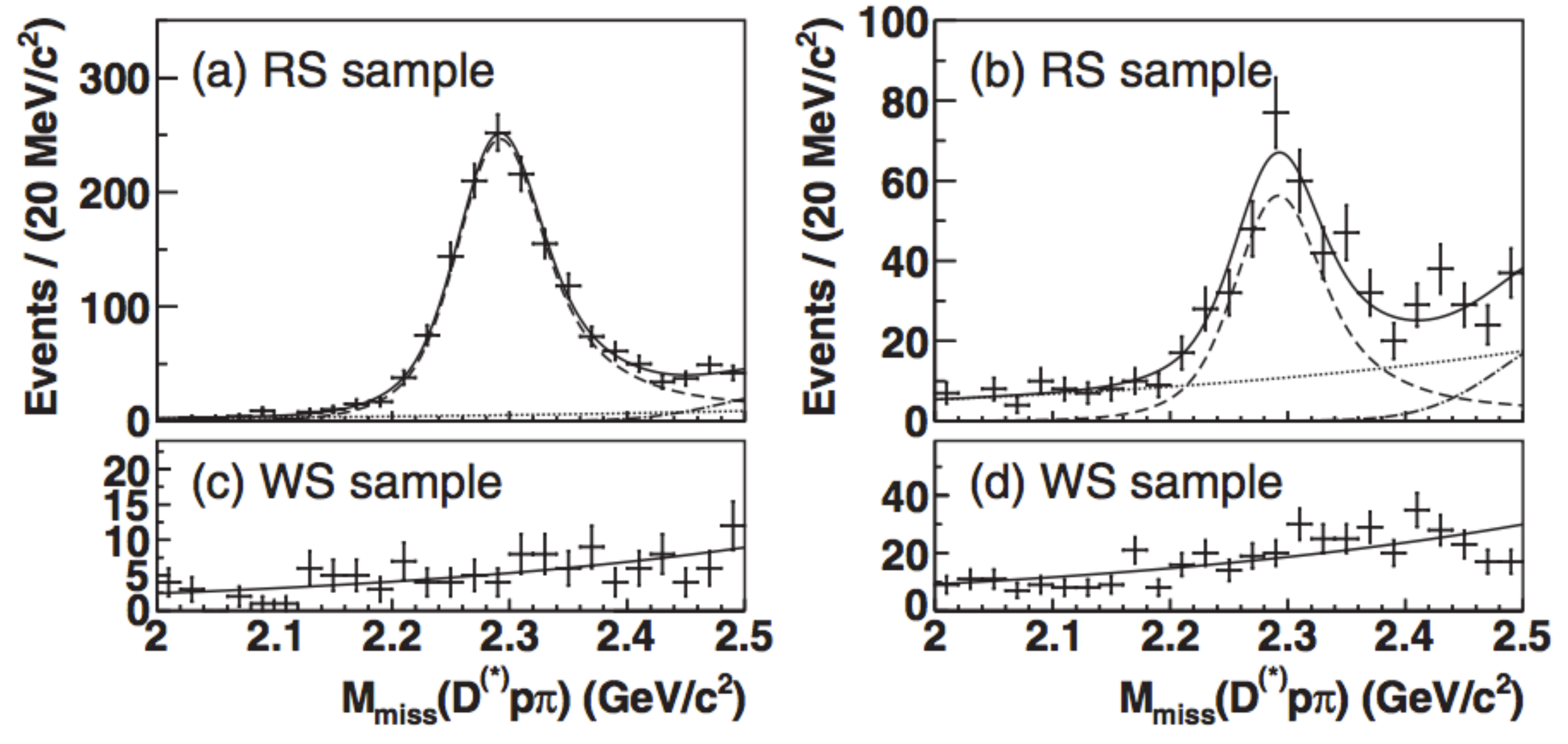}
\figcaption{\label{fig1}The $M_{miss}(D^{(*)}\bar{p}\pi)$ data distributions of exclusively reconstructed $\Lambda_c^+$ candidates. (a) and (c) for the SR region and (b) and (d) SB region of $M(pK\pi)$ for the RS and WS samples, respectively, with superimposed fit results (solid line).}
\end{center}

\acknowledgments{We thank Belle collaboration for the contribution for KEKB in achieving the highest luminosity
and most stable machine condition and we are grateful to K. Thomas, M. Nakao, N. K. Nisar, C. P. Shen and  V. Bhardwaj for advices and help.}

\end{multicols}

%\vspace{10mm}

\vspace{-1mm}
\centerline{\rule{80mm}{0.1pt}}
\vspace{2mm}

\begin{multicols}{2}

\end{multicols}

\clearpage

\end{CJK*}

\begin{thebibliography}{90}

\vspace{3mm}
\bibitem{Belle} A. Abashian {\it et al.} (Belle Collaboration), Nucl. Instrum. Methods Phys. Res., Sect. A {\bf 479}, 117 (2002); also see the detector section in J. Brodzicka et al., Prog. Theor. Exp. Phys. {\bf 2012}, 04D001 (2012).

\bibitem{obs_LHCb} R. Aaij {\it et al.} (LHCb Collaboration), Phys. Rev. Lett. {\bf 110}, 101802 (2013); Phys. Rev. Lett. {\bf 111}, 251801 (2013).

\bibitem{obs_CDF} T. Aaltonen {\it et al.} (CDF Collaboration), Phys. Rev. Lett. {\bf 111}, 231802 (2013).

\bibitem{lab1} B. R. Ko {\it et al.} (Belle Collaboration), Phys. Rev. Lett. {\bf 112}, 111801 (2014).

\bibitem{HFAG} Heavy Flavor Averaging Group (HFAG): Charm Physics Paramters, \url{http://www.slac.stanford.edu/xorg/hfag/charm}

\bibitem{lab2} T. Peng {\it et al.} (Belle Collaboration), Phys. Rev. D {\bf 89}, 091103(R) (2014). 

\bibitem{lab3} M. Staric {\it et al.} (Belle Collaboration), arXiv:1509.08266, submitted to Phys. Lett. B; Belle note 1233 v2. 

\bibitem{lab4} N. K. Nisar  {\it et al.} (Belle Collaboration), Phys. Rev. Lett. {\bf 112}, 211601 (2014).

\bibitem{lab5} N. K. Nisar {\it et al.} (Belle Collaboration), Belle note 1370 v11.

\bibitem{PDG2014} K. A. Olive {\it et al.} (Particle Data Group), Chin. Phys. C, {\bf 38}, 090001 (2014).

\bibitem{lab6} Y. Kato {\it et al.} (Belle Collaboration), Phys. Rev. D {\bf 89}, 052003 (2014).

\bibitem{lab7} S. H. Lee {\it et al.} (Belle Collaboration), Phys. Rev. D {\bf 89}, 091102(R) (2014).

\bibitem{lab8} A. Zupanc {\it et al.} (Belle Collaboration), Phys. Rev. Lett. {\bf 113}, 042002 (2014).

\end{thebibliography}
\end{document}